%% file: turkey.tex
\documentstyle[lprocldpf,epsf]{article}

\input{defs}

\def\Journal#1#2#3#4{{#1} {\bf #2}, #3 (#4)}

\def\PRL{\em Phys. Rev. Lett.}
\def\PRB{{\em Phys. Rev.} B}
 
\def\p{{\bf p}}
\def\q{{\bf q}}
\def\QQ{{\bf Q}}
\def\om{\omega}

\def\gtwid{\mathrel{\raise.3ex\hbox{$>$\kern-.75em\lower1ex\hbox{$\sim$}}}}
\def\ltwid{\mathrel{\raise.3ex\hbox{$<$\kern-.75em\lower1ex\hbox{$\sim$}}}}

\begin{document}

\title{PAIRING MECHANISM IN TWO HUBBARD MODELS THAT SHOW \\
{$d_{x^2-y^2}$} PAIRING
}

\author{D.J. SCALAPINO}

\address{Department of Physics, University of California,
Santa Barbara,\\ CA 93106--9530, USA}


\maketitle\abstracts{
Here we examine various aspects of the pairing mechanism for two models,
the two-dimensional and two-leg ladder Hubbard models.
Both of these models exhibit pairing correlations with $\dxy$ symmetry.
However, the undoped insulating states of these two systems are
different with the two-dimensional lattice characterized by a ground
state with long-range antiferromagnetic order and the two-leg ladder
having a spin gap in its ground state.
Our aim is to gain a better understanding of the underlying pairing
mechanism which causes $\dxy$ pairing in these two models.
}

\section{Introduction}

Understanding the pairing mechanism responsible for superconductivity in
the high-temperature superconducting cuprates remains one of the central
issues in condensed matter physics.
Various experiments~\cite{Sch,Scal,Van}
strongly suggest that the dominant symmetry of the
gap in these materials is $\dxy$.
However, the implications of this for the pairing mechanism remain
controversial~\cite{Stan}.
In one view~\cite{Bickers,Moriya,Pines}, 
antiferromagnetic spin-fluctuation exchange models provide
a simple framework for understanding the $\dxy$ symmetry of the gap.
In addition, various phenomenological calculations based upon these
ideas have often provided remarkable fits to data~\cite{Scal}.
However, 
the effect of vertex corrections~\cite{Sch1} as well as
the unusual normal state properties and in particular the
pseudogap phenomena~\cite{Houston}
observed in the under-doped materials raise key
questions which suggest that important pieces of the puzzle remain to be
understood.
Thus it seems useful to review what is understood about systems which
show $\dxy$ pairing fluctuations.
Specifically, we propose to review some results which have been obtained
for the two-dimensional Hubbard model and the two-leg Hubbard ladder.
While it is not known whether the doped two-dimensional Hubbard model
has a low-temperature superconducting phase, there is numerical evidence
for $\dxy$ pairing fluctuations in this 
model~\cite{Bickers,Bul1,Bul2}.
For the doped two-leg ladder, groundstate calculations~\cite{Noack}
clearly show the
presence of $\dxy$-like pairs.
Furthermore, the undoped insulating states for these two systems are
different.
The ground state of the two-dimensional Hubbard model has long-range
antiferromagnetic order~\cite{Hirsch}, 
while that of the two-leg Hubbard ladder is
characterized by a spin gap~\cite{Noack}.
Both, however, have strong short-range antiferromagnetic correlations.
Thus it is of interest to compare them and
here we will examine these two cases with the aim of gaining insight
into the mechanism underlying the tendency towards $\dxy$ pairing in
these systems.

After a summary of the energy scales which arise in the Hubbard
model, we first turn to a discussion of the two-dimensional Hubbard
model.
Here we note the development of short-range antiferromagnetic
correlations in the doped two-dimensional Hubbard model as the
temperature is lowered below the exchange interaction $4\,t^2/U$.
We examine the effect of these correlations on the structure of the
quasiparticle spectrum and the effective particle-particle interaction.
We then discuss how this interaction leads to a tendency towards $\dxy$
pairing.

Following this, we turn to the example of the two-leg Hubbard ladder.
Here we first discuss the half-filled insulating case, distinguishing
between a band insulator and a strongly correlated spin gap insulator.
Then we study the case of a doped ladder and discuss the nature of the
observed pairing correlations.

The Hubbard model provides a simple model of the CuO$_2$ system:
\begin{equation}
H=-t\sum_{\langle ij\rangle ,s} \left( c^\dagger_{is}c_{js} +
c^\dagger_{js}c_{is} \right) + U \sum_i n_{i\uparrow}n_{i\downarrow}.
\label{eq:ham}
\end{equation}
Here $c^\dagger_{is}$ creates an electron of spin $s$ on site $i$, in
the first term the sum is over near-neighbor sites, and $n_{is} =
c^\dagger_{is}c_{is}$ is the occupation number for 
electrons with spin $s$ on site $i$.
The one-electron transfer between near-neighbor sites is $t$, and $U$ is
an onsite Coulomb energy.
In Eq.~(1), we have chosen to keep only a near-neighbor hopping, but one
could of course also add a next-near-neighbor hopping $t'$.

The bare energy scale is set by the bandwidth $8\,t$ and the effective
Coulomb interaction $U$, which are both of order 
several electron volts.
Near half-filling, electrons on neighboring sites tend to align
antiferromagnetically so as to lower their energy by the exchange
interaction $J=4\,t^2/U$.
This interaction is of order a tenth of an electron volt and, as we will
see, sets the energy, or temperature scale, below which
antiferromagnetic correlations, the low-energy structure in the
single-particle spectral weight, and the pairing interaction develop.

While Monte Carlo~\cite{Dag} and Lanczos calculations~\cite{Par}
for a $4\times4$ lattice find that two holes
added to the half-filled Hubbard ground state form a $\dxy$ bound state,
and density matrix renormalization group calculations~\cite{Noack} 
find that
$\dxy$-like pairs are formed on two-leg Hubbard ladders, it is not known
what happens for the two-dimensional Hubbard model.
It is possible that on an energy scale of order $J/10$, a $\dxy$
superconducting state forms.
However, this may well require modifications of the model, such as an
additional near-neighbor 
$\Delta J {\bf S}_i \cdot{\bf S}_j$ 
term or possibly a next-near-neighbor hopping $t'$.
Nevertheless, it is known that as the temperature is reduced below $J$,
$\dxy$ pairing correlations develop in the doped two-dimensional Hubbard
model, and here we will examine why this happens.

\section{Two-Dimensional Hubbard Lattice}

\begin{figure} 
\centerline{\epsfysize=6.8cm \epsffile[-30 184 544 598] {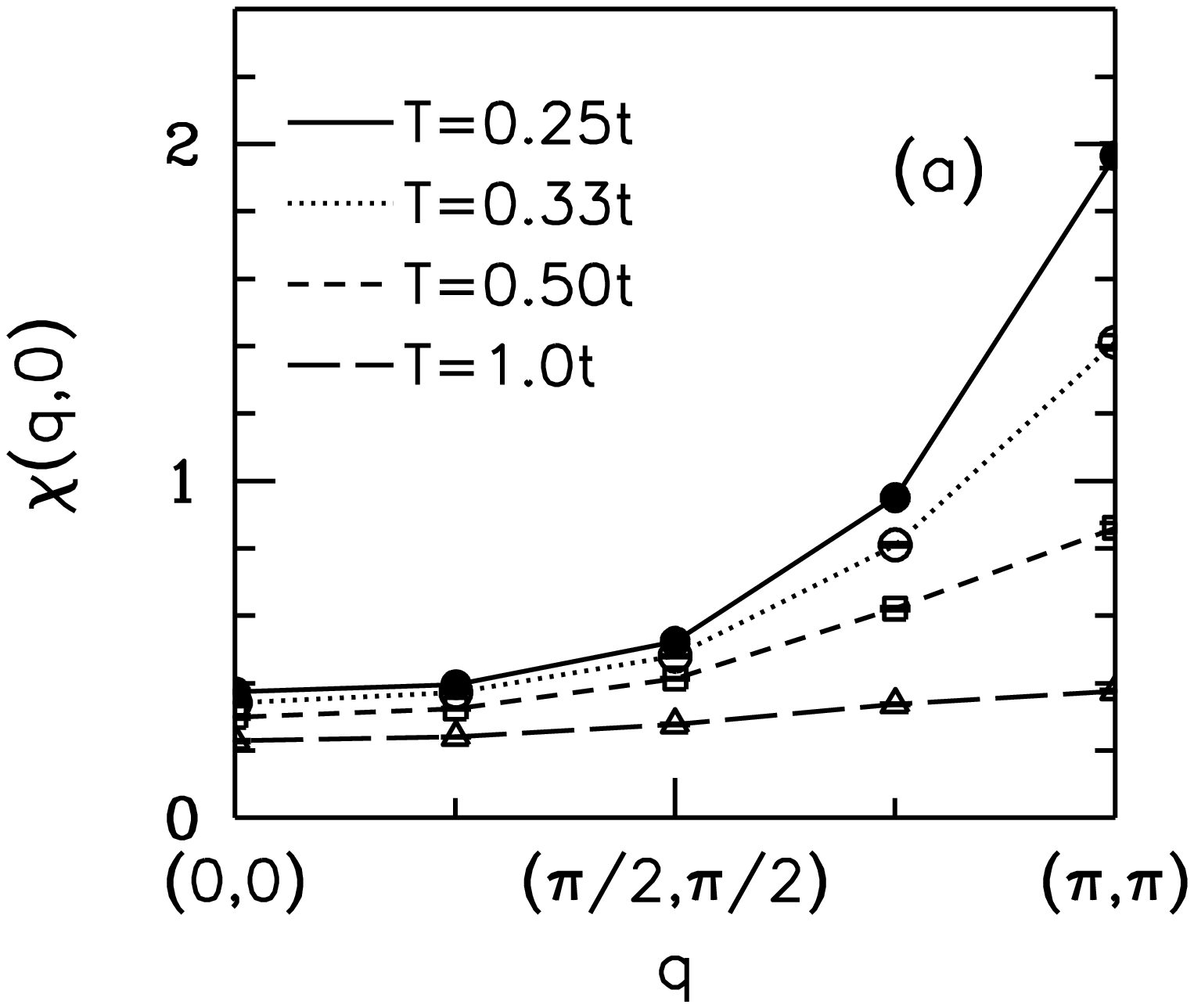}
\epsfysize=6.8cm \epsffile[98 184 672 598] {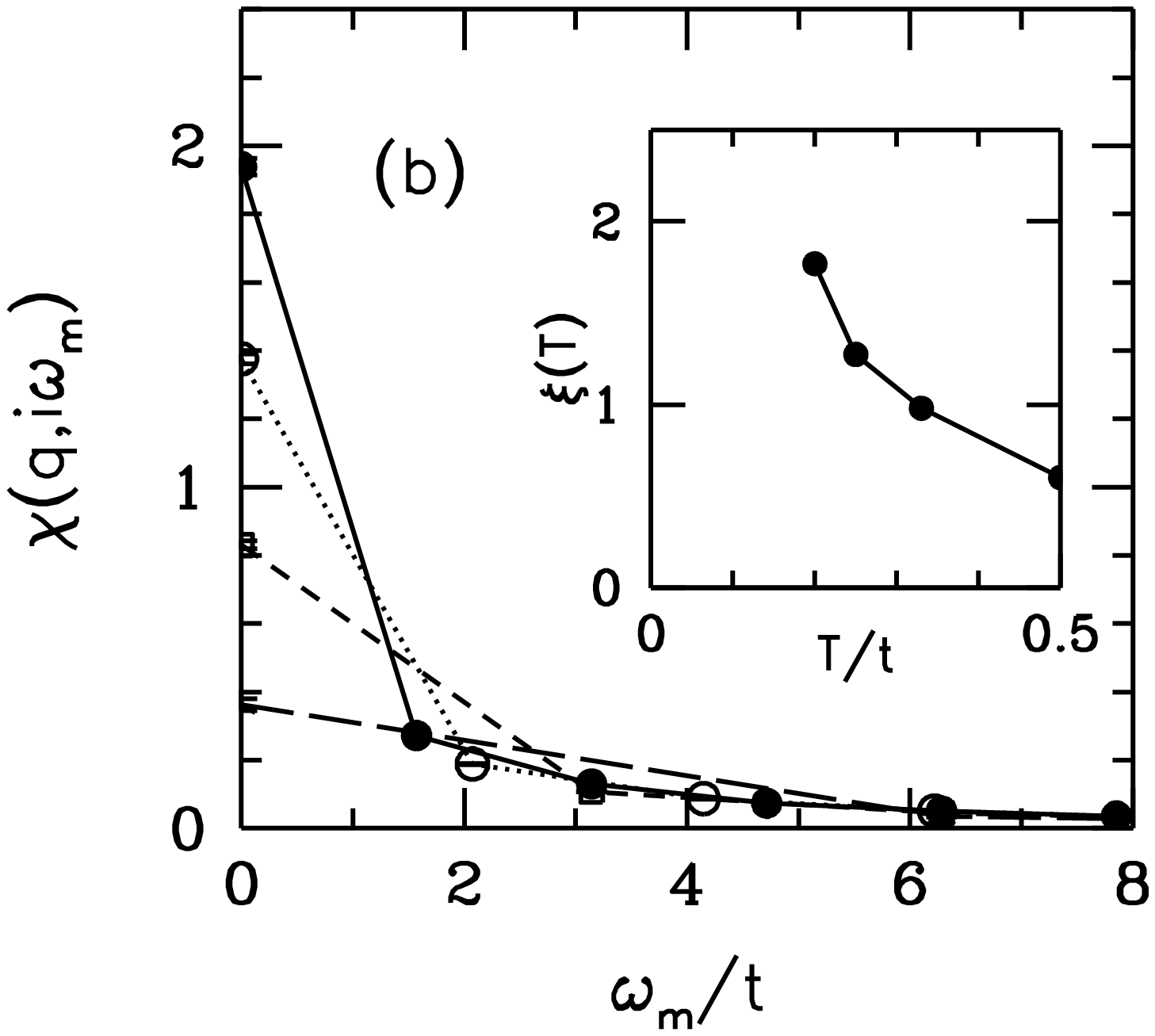}}
\vspace{0.3cm}
\caption{
(a) Magnetic susceptibility $\chi({\bf q},0)$ versus
{\bf q} along the (1,1) direction for various temperatures.
These results are for an $8\times8$ lattice with $U/t=4$ and a filling
$\nang=0.875$.
(b) Matsubara frequency dependence of $\chi({\bf q},i\omega_m)$ for
${\bf q}=\pp$ at the same temperatures as in Fig. 1(a).
Note that as $T$ decreases below $4\,t^2/U$, short-range
antiferromagnetic fluctuations develop.
Inset: Temperature dependence of the antiferromagnetic 
correlation length $\xi$.
\label{fig:chiqw}}
\end{figure}

\begin{figure} 
\centerline{\epsfysize=6.8cm \epsffile[-30 184 544 598] {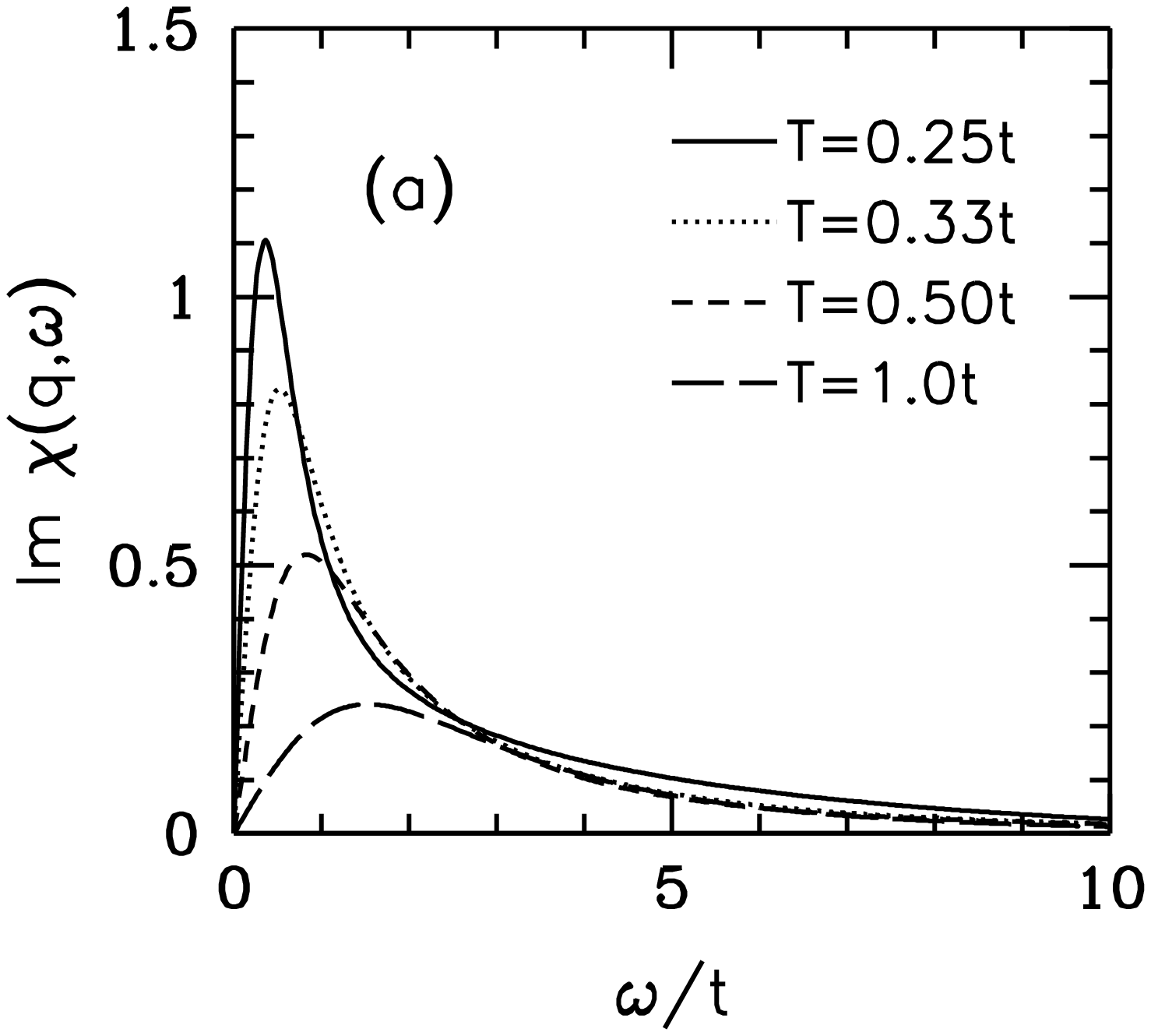}
\epsfysize=6.8cm \epsffile[98 184 672 598] {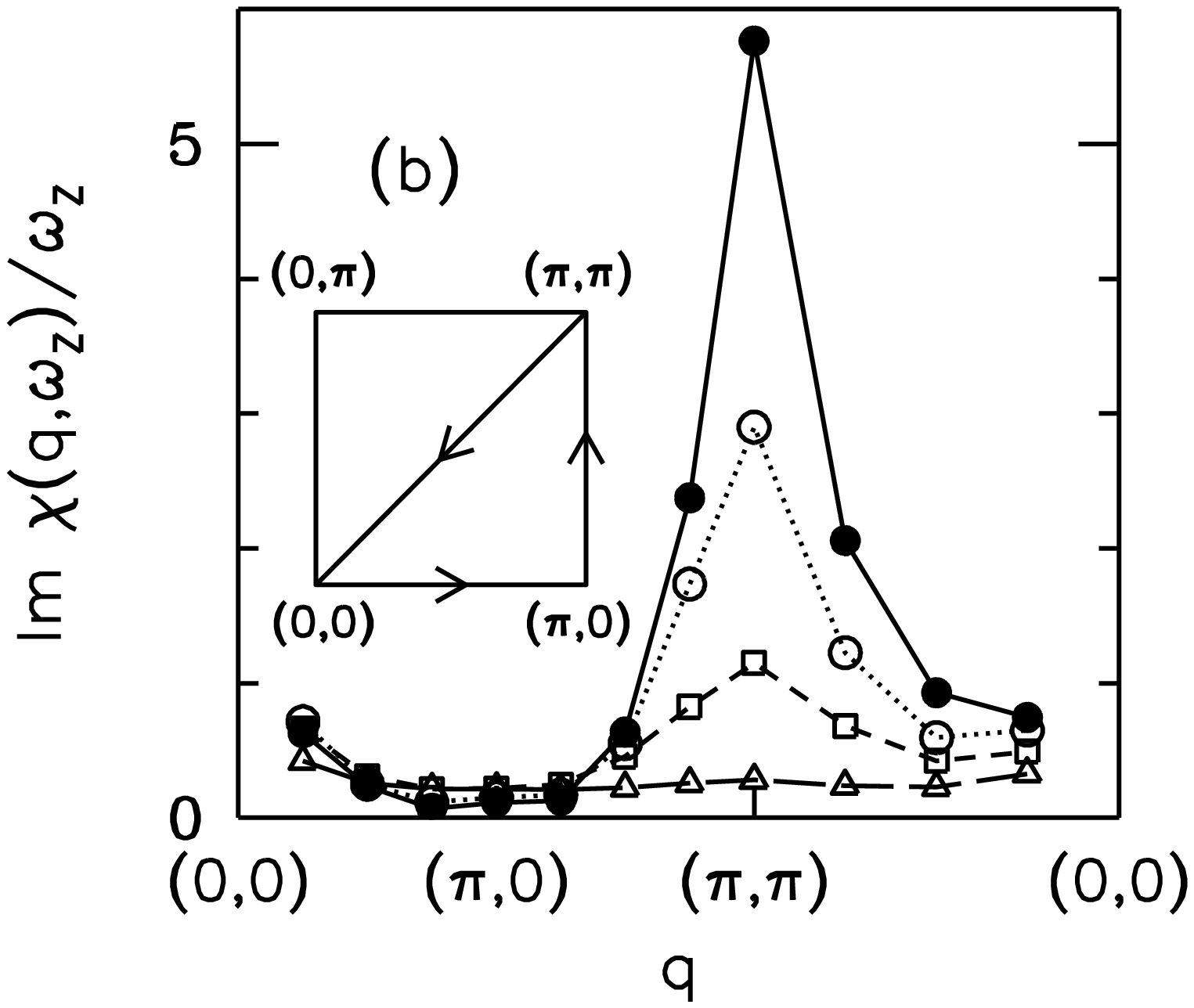}}
\vspace{0.3cm}
\caption{
(a) Spin--fluctuation spectral weight 
${\rm Im}\,\chi(\q,\om)$ versus $\om$ at $\q=(\pi,\pi)$
for various temperatures.
(b) ${\rm Im}\,\chi(\q,\om_Z)/\om_Z$ versus $\q$ in the limit of
$\om_Z\rightarrow 0$ for the same temperatures as in Fig. 2(a).  
Here $\q$ is plotted along the path shown in the inset.  
These figures are for $U/t=4$ and $\nang=0.875$.
\label{fig:imchi}}
\end{figure}

At half-filling, $\langle n_{i\uparrow} + n_{i\downarrow}\rangle = 1$,
the 2D Hubbard model develops long-range antiferromagnetic order as the
temperature goes to zero~\cite{Hirsch}.
In the doped case, strong short-range antiferromagnetic correlations
develop as the temperature decreases below $J$.
This is clearly seen in the temperature dependence of the wave vector
and Matsubara frequency-dependent magnetic susceptibility
\begin{equation}
\chi({\bf q},i\omega_m) = {1\over N} \sum_{\lb}
  \int^\beta_0 d\tau\, 
   e^{i\omega_m\tau} 
   e^{-i\q\cdot\lb} 
     \left\langle m^-_{i+\ell}(\tau) m^+_{i}(0) \right\rangle.
\label{eq:chiqw}
\end{equation}
Here 
$m^+_i(0) = c^{\dagger}_{i\uparrow}c_{i\downarrow}$
and 
$m^-_{i+\ell}(\tau) = e^{H\tau} m^-_{i+\ell}(0) e^{-H\tau}$,
where $m^-_{i+\ell}(0)$ is the hermitian conjugate of 
$m^+_{i+\ell}(0)$.
Monte Carlo results for $\chi({\bf q},0)$ versus $\q$ along the $(1,1)$
axis for an $8\times8$ lattice with $U/t=4$ and a filling $\nang=0.875$
are shown in Fig.~\ref{fig:chiqw}(a).
In Fig.~\ref{fig:chiqw}(b), the Matsubara frequency dependence of 
$\chi({\bf q},i\omega_m)$ versus 
$\omega_m=2\,m\pi T$ is shown for ${\bf q} = \pp$.
The inset of Fig.~\ref{fig:chiqw}(b) shows the temperature dependence of the 
antiferromagnetic correlation length $\xi$.
Here, $\xi^{-1}$ is defined as the half--width at half--maximum of 
$\chi(\q,0)$.
The low--frequency nature of the antiferromagnetic correlations are
seen in Fig. \ref{fig:imchi}(a) where 
${\rm Im}\,\chi(\q=(\pi,\pi),\om)$ versus $\om$ is plotted 
at various temperatures.
In addition, Fig. \ref{fig:imchi}(b) shows
${\rm Im}\,\chi(\q,\om_Z)/\om_Z$ versus $\q$ in the 
$\om_Z\rightarrow 0$ limit.
This is the quantity which determines the NMR $T_1^{-1}$
response of the system.
These figures clearly show the development of 
significant short--range and low--frequency antiferromagnetic correlations 
as the temperature decreases below 
$J\simeq 4\,t^2/U$.

\begin{figure} 
\centerline{\epsfysize=6.8cm \epsffile[-30 184 544 598] {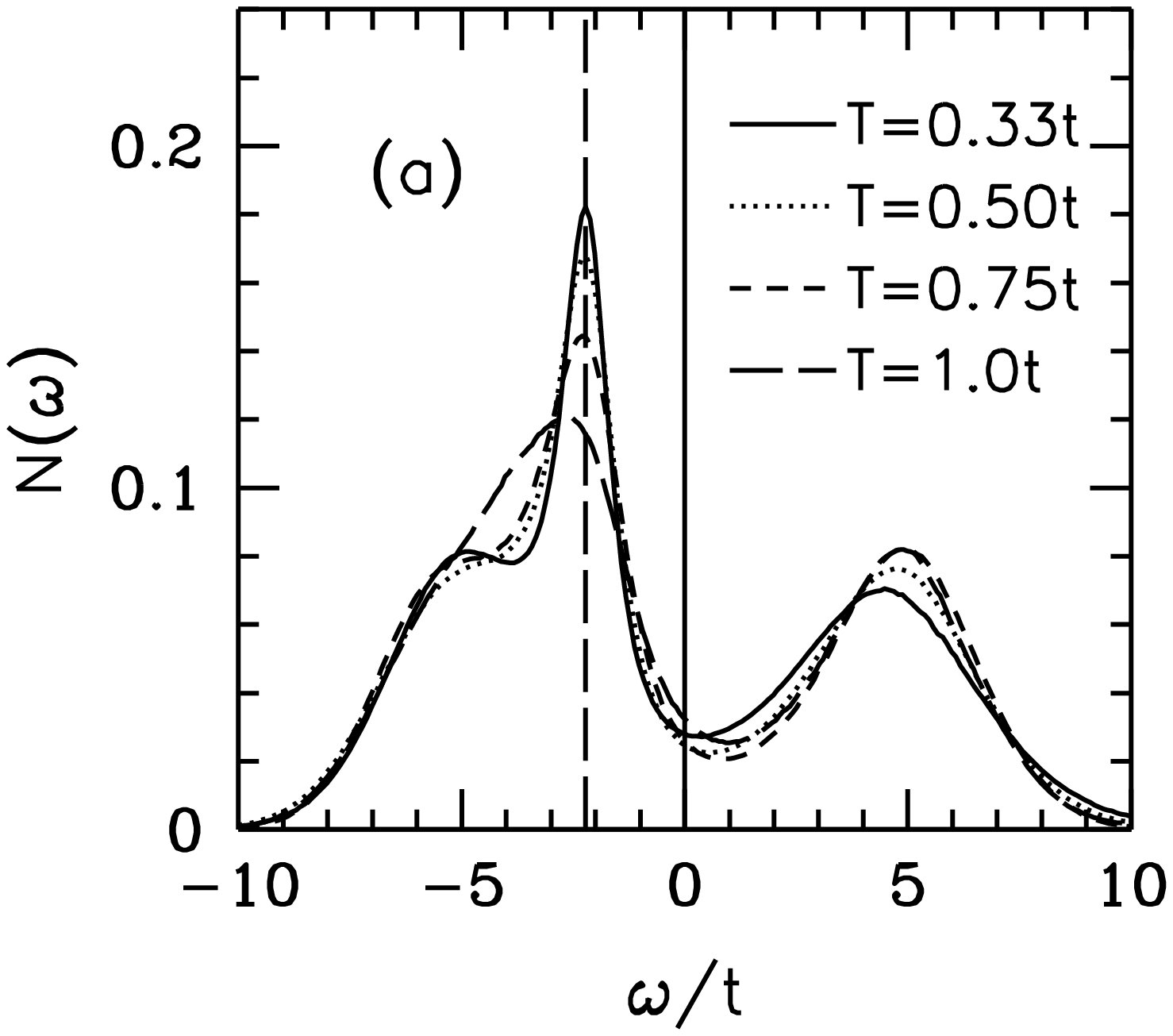}
\epsfysize=6.8cm \epsffile[98 184 672 598] {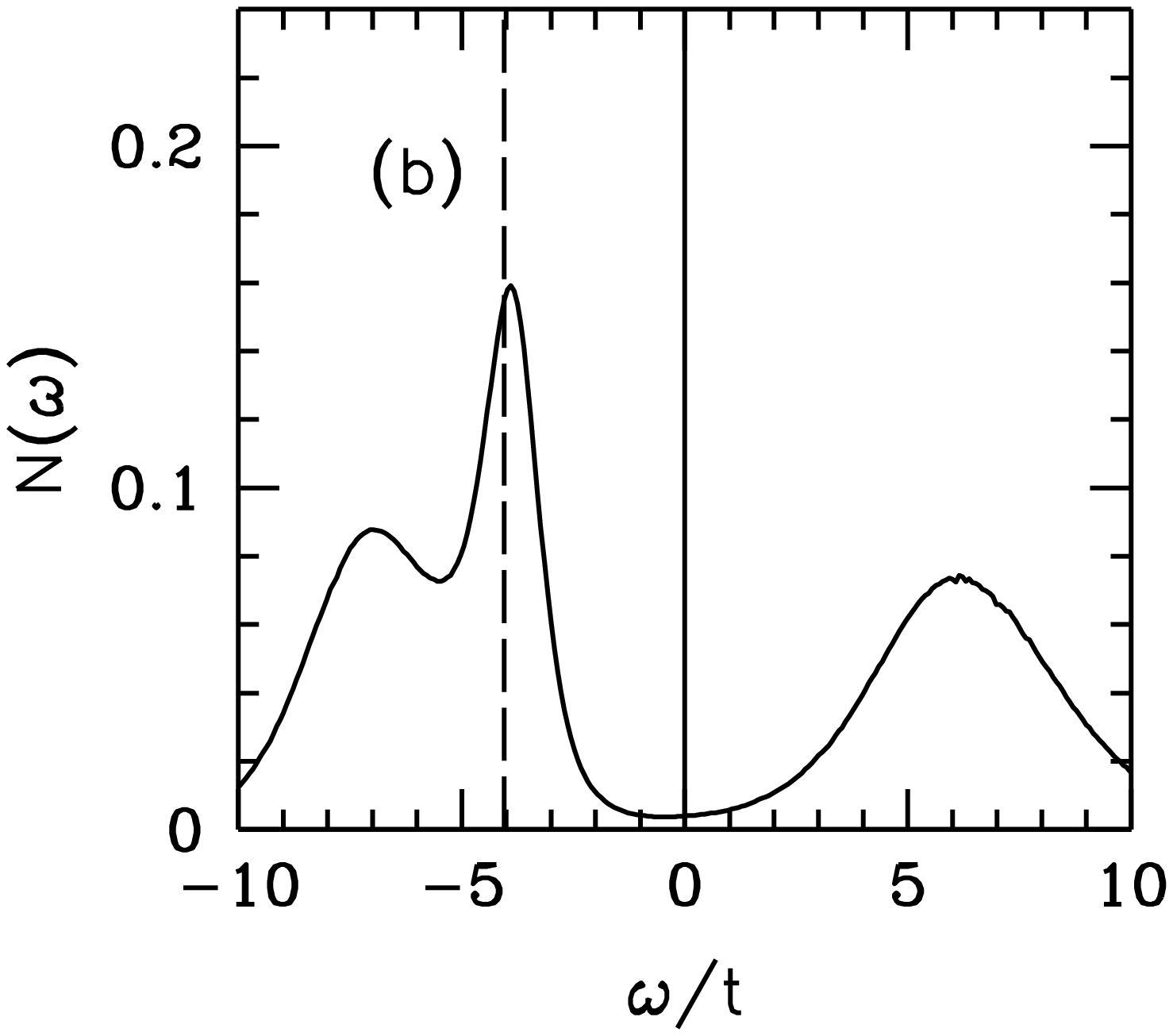}}
\vspace{0.3cm}
\caption{
(a) Evolution of the single--particle density of states 
$N(\om)$ with temperature for $U/t=8$ and $\nang=0.875$.
(b) $N(\om)$ versus $\om$ for $U/t=12$, $\nang=0.875$ and
$T=0.5t$.
The vertical dashed lines denote the chemical potential.
\label{fig:Nw}}
\end{figure}

\begin{figure} 
\centerline{\epsfysize=7.5cm \epsffile[-207 144 367 718] {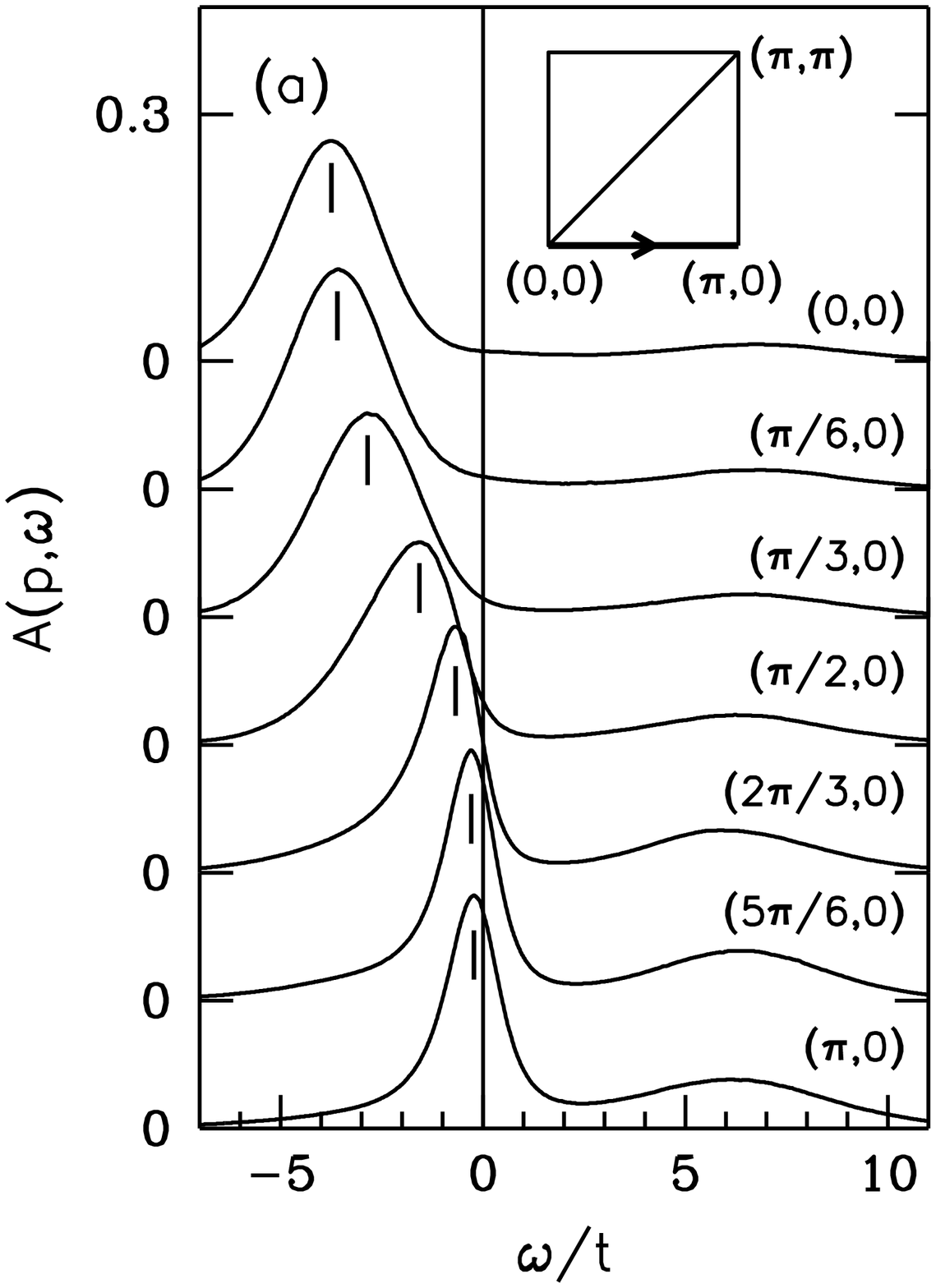}
\epsfysize=7.5cm \epsffile[18 144 592 718] {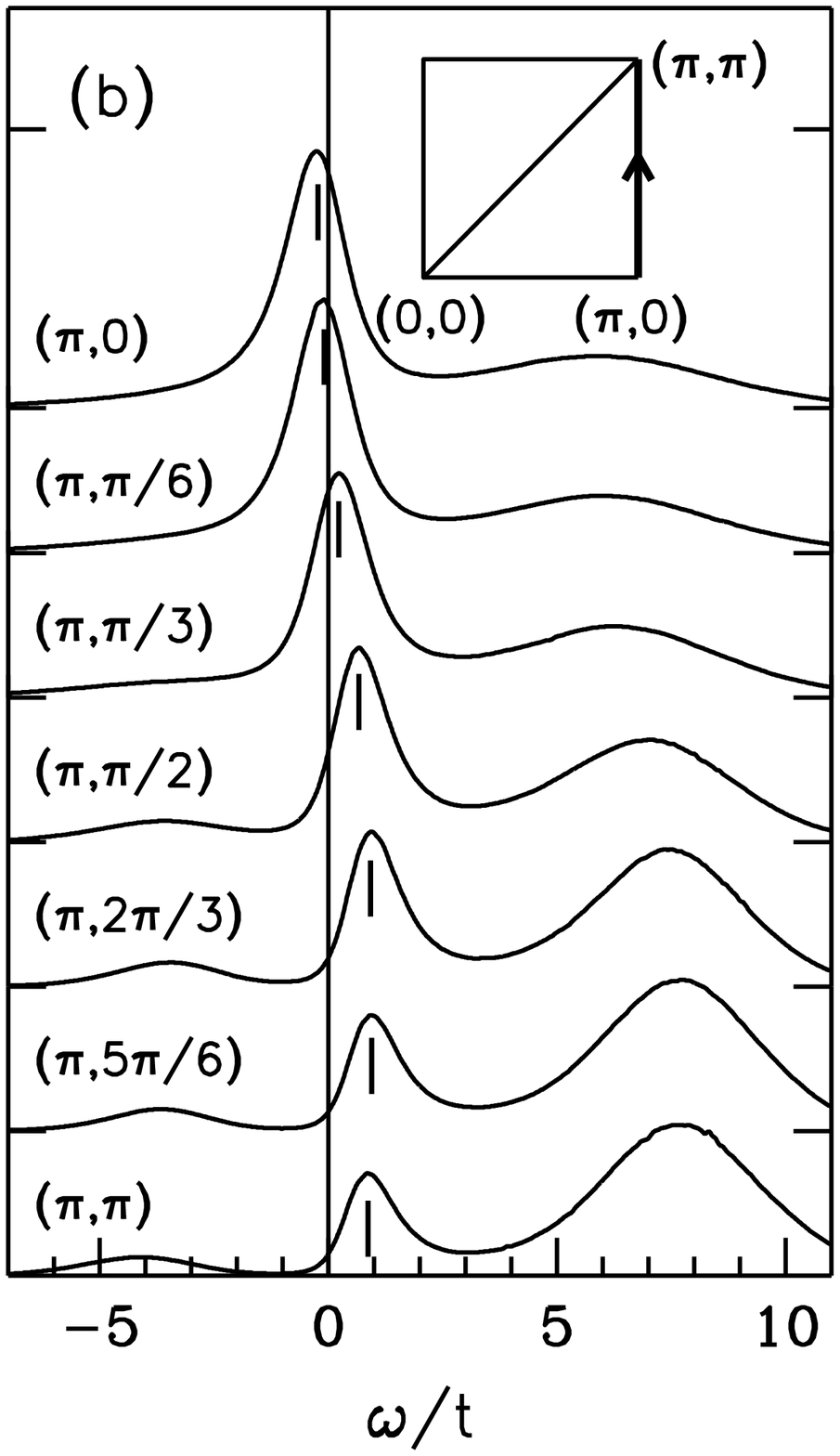}
\epsfysize=7.5cm \epsffile[243 144 817 718] {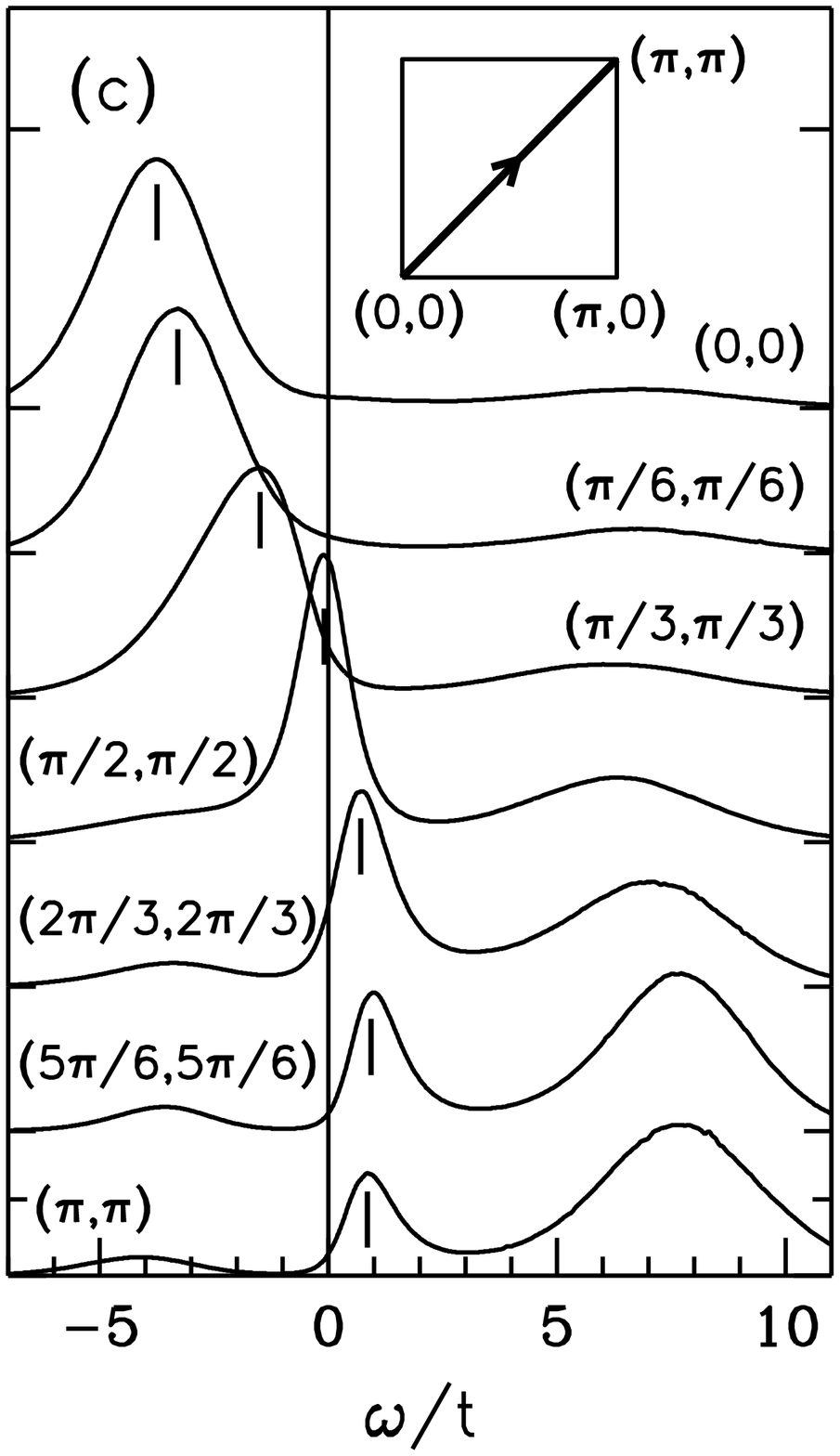}}
\caption{
Single--particle spectral weight along various cuts 
in the Brillouin zone is shown for $U/t=8$ and $\nang=0.875$ on a $12\times12$
lattice at $T=0.5\,t$.
\label{fig:Apw}}
\end{figure}

\begin{figure}
\centerline{\epsfysize=8cm \epsffile[18 184 592 598] {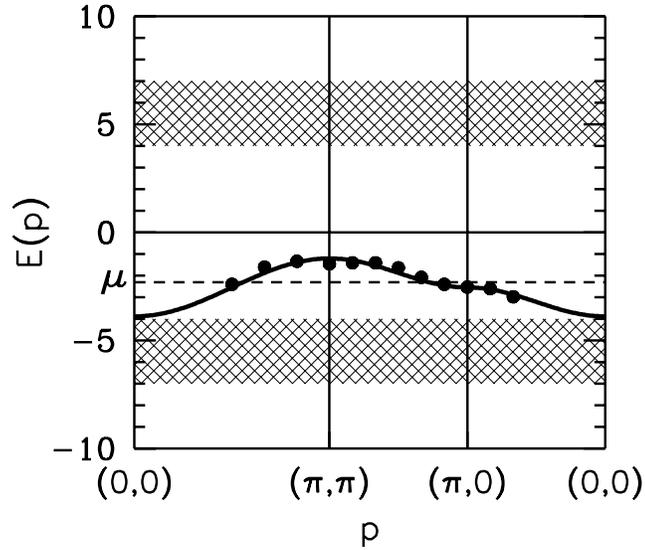}}
\vspace{0.3cm}
\caption{
Dispersion of the quasiparticle peak in the spectral
weight versus {\bf p}
for $U/t=8$, $\nang=0.875$ and $T=0.5t$.
The solid points mark the low--energy peaks of $A(\p,\omega)$ shown in
Fig.~4, 
and the solid curve represents an estimate of the quasiparticle
dispersion using these data and Lanczos results for {\bf p} near (0,0).
The broad darkened areas represent the incoherent spectral weight in the
upper and lower Hubbard bands.
The horizontal dashed line denotes the chemical potential $\mu$.
\label{fig:Ep}}
\end{figure}

As these antiferromagnetic correlations develop, the single-particle
spectral weight
\begin{equation}
A({\bf p},\omega) = -{1\over \pi}\,\im\,G({\bf p},
i\omega_n\to\omega+i\delta) 
\label{eq:apw}
\end{equation}
and the density of states
\begin{equation}
N(\omega) = {1\over N}
\sum_{\p} A({\bf p},\omega)
\label{eq:nw}
\end{equation}
also change.
Figure \ref{fig:Nw}(a) shows 
the temperature evolution of 
$N(\omega)$ for $U/t=8$ and $\nang=0.875$.
Figure \ref{fig:Nw}(b) shows $N(\om)$ for 
$U/t=12$ and $T=0.5t$.
As the temperature is lowered, a peak appears on the upper edge of the
lower Hubbard band.
This peak arises from a narrow quasiparticle band shown in the
single-particle spectral weight $A({\bf p},\omega)$ of 
Fig.~\ref{fig:Apw} and
plotted as the solid curve in Fig.~\ref{fig:Ep}.
As the momentum {\bf p} goes towards the $\Gamma$ point $(0,0)$, we
believe that the quasiparticle peak is obscured by the lower Hubbard
band because of the resolution of the maximum entropy technique which we
have used.
Indeed, at the $\Gamma$ point a separate quasiparticle peak is found
from Lanczos exact diagonalization~\cite{Lancapw}
on a $4\times4$ lattice.
The solid line in Fig.~\ref{fig:Ep} shows what we believe is the 
quasiparticle dispersion relation.
As clearly evident in the spectral weight shown 
in Figs.~\ref{fig:Apw}(a) and (b), the 
quasiparticle dispersion is anomalously flat near the $(\pi,0)$
corner.
As discussed by various authors~\cite{Bul3,Moreo}, 
this reflects the influence of the
antiferromagnetic correlations on the quasiparticle excitation energy.
It is clear that the peak structure in $N(\omega)$ also arises from the
short-range antiferromagnetic correlations and is a many-body effect
rather than simply a non-interacting band Van Hove singularity.

\begin{figure} 
\begin{picture}(30,15)
\end{picture}
\includegraphics{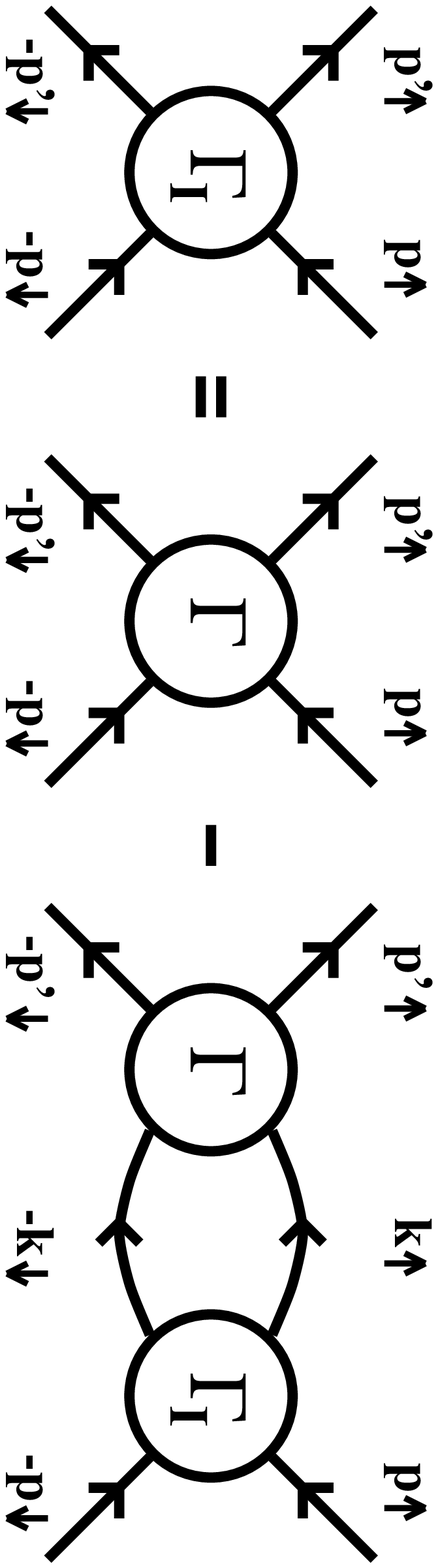}
\includegraphics{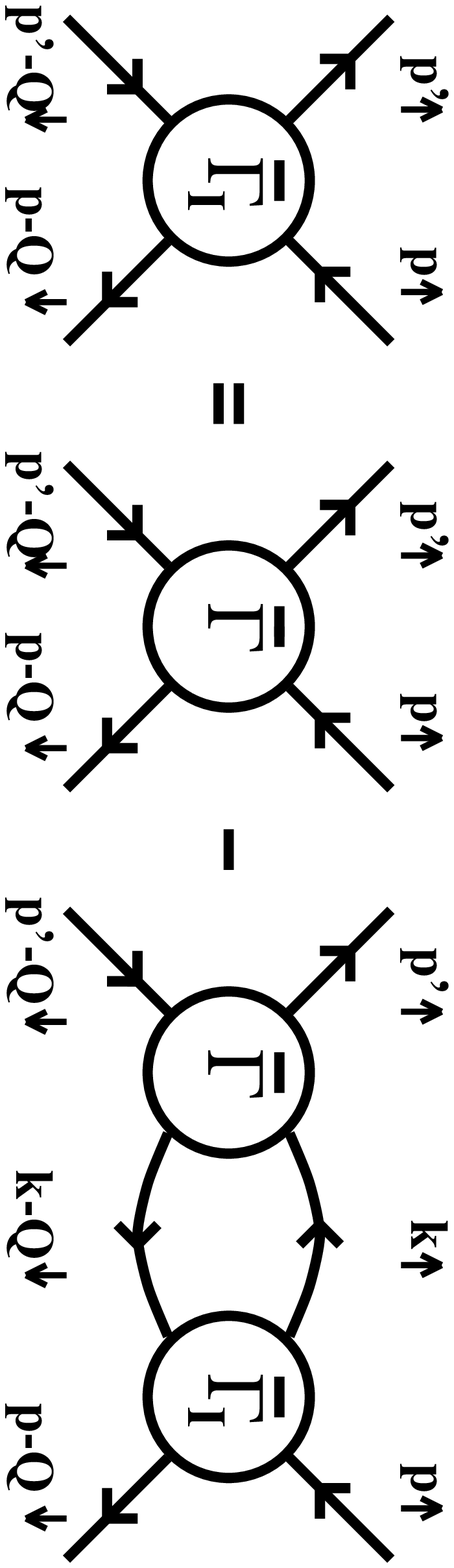}
\vspace{6.5cm}
\caption{Particle-particle and particle-hole $t$-matrix equations.
\label{fig:pp}}
\end{figure}

\begin{figure}
\centerline{\epsfysize=6.8cm \epsffile[-30 184 544 598] {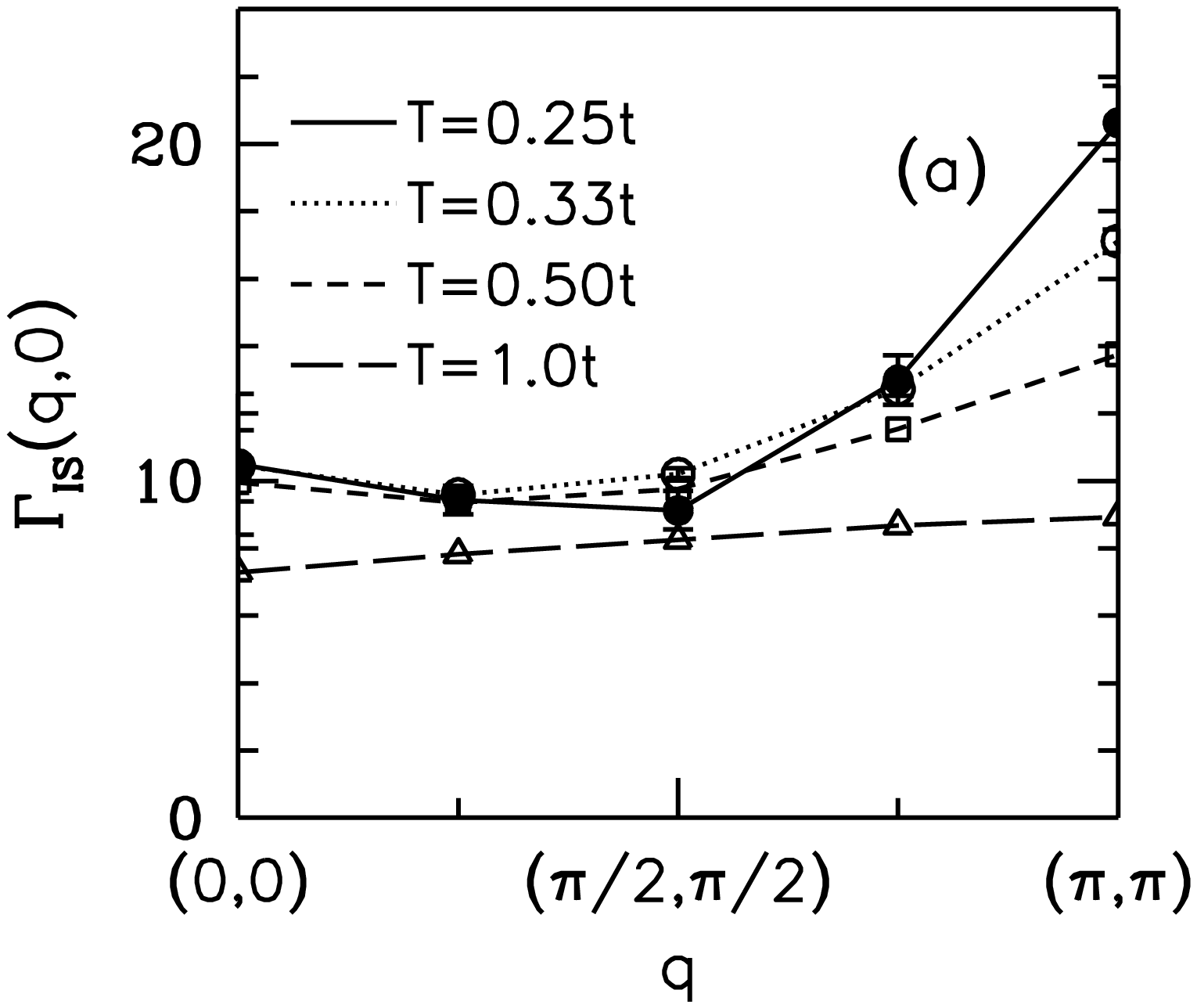}
\epsfysize=6.8cm \epsffile[98 184 672 598] {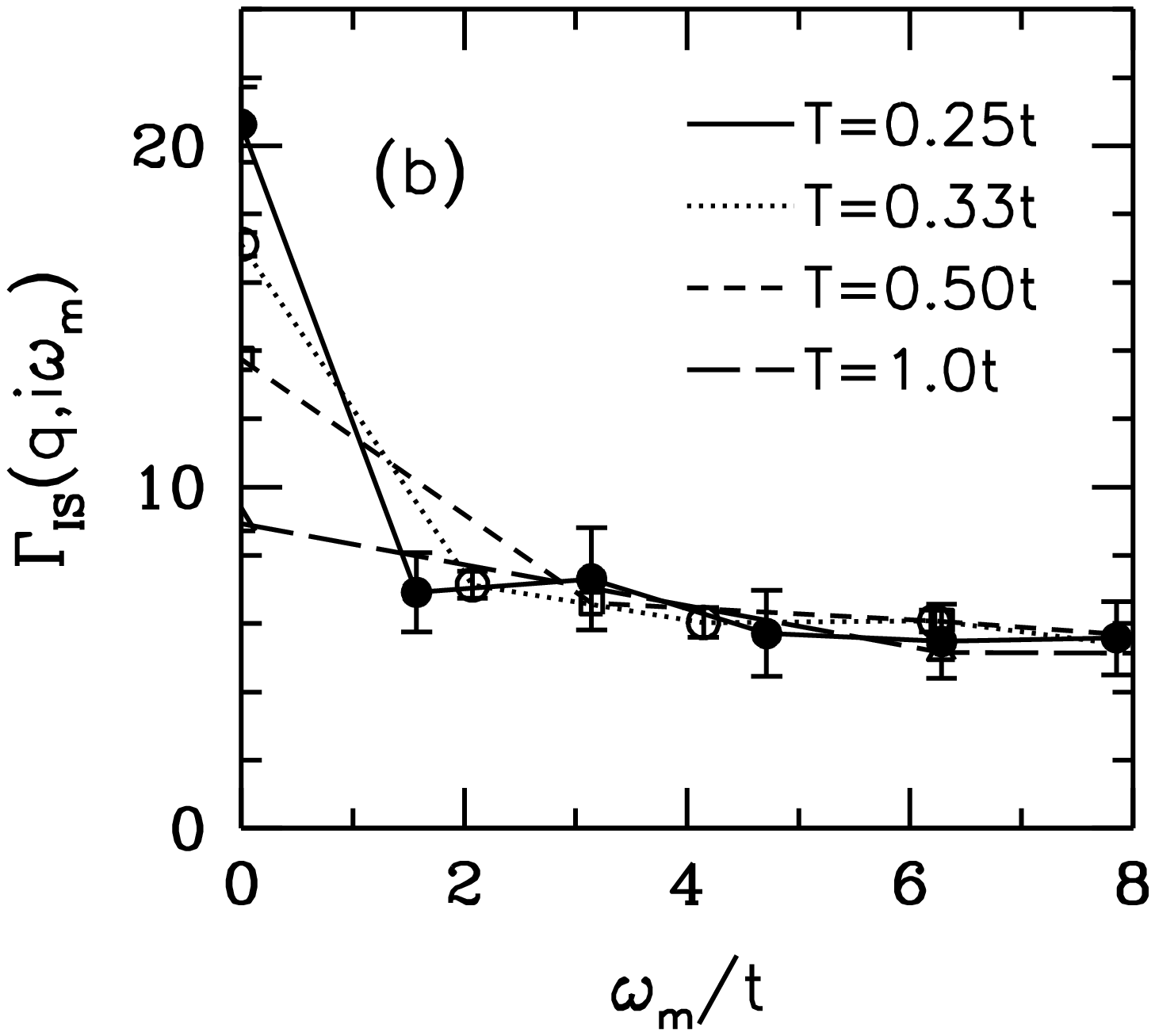}}
\vspace{0.3cm}
\caption{
(a) Singlet irreducible particle--particle vertex for zero
energy transfer $\Gamma_{\rm IS}({\bf q},i\omega_m=0)$ 
versus ${\bf q}$ along the (1,1) direction.
Here $U/t=4$ and $\nang=0.875$.
As the temperature decreases below $J=4\,t^2/U$, the strength of the
interaction is enhanced at large momentum transfer.
(b) Energy transfer dependence of 
$\Gamma_{\rm IS}({\bf q},i\omega_m)$ for ${\bf q} = \pp$, 
$\omega_n=\pi T$, and $\omega_{n'} =
\omega_n+\omega_m$ is shown for various temperatures.
Again note the similarity to $\chi({\bf q},i\omega_m)$ in Fig.~1.
\label{fig:gamI}}
\end{figure}


Monte Carlo calculations have also been used to determine the
two-fermion scattering vertex $\Gamma(p_1,p_2,p_3,p_4)$.
Here $p_1$ stands for $({\bf p}_1,i\omega_{n_1})$ and $\sigma$.
Using $\Gamma$ and the single-particle Green's function $G(p)$, one can
solve the $t$-matrix equations shown diagrammatically in 
Fig.~\ref{fig:pp} to
obtain the irreducible particle-hole and particle-particle vertices.
The singlet irreducible particle-particle vertex $\Gamma_{\rm IS} (p',
-p', p, -p)$ in the zero center-of-mass momentum and energy channel
represents the effective pairing interaction.
In Fig.~\ref{fig:gamI}(a), 
$\Gamma_{\rm IS}(q=p-p')$ is plotted for {\bf q} along the
$(1,1)$ direction and $i\omega_n= i\omega_{n'} = i\pi T$, corresponding
to zero Matsubara energy transfer.
The Matsubara frequency dependence 
of this vertex for energy transfer
$\omega_m=\omega_{n'} - \omega_n$ is shown in 
Fig.~\ref{fig:gamI}(b) for ${\bf q}=\pp$.
Comparing Figs.~\ref{fig:chiqw} and \ref{fig:gamI}, 
one clearly sees that the {\bf q} and
$\omega_m$ structure of the interaction and $\chi(\q,i\omega_m)$ are
similar, both reflecting the development of the antiferromagnetic
correlations as $T$ is reduced below $J$.

Given the Monte Carlo results for the irreducible particle-particle
vertex $\Gamma_{\rm IS}(p',-p',p,-p)$ in the zero energy and
center-of-mass momentum channel, which we will denote by $\Gamma_{\rm
IS}(p'|p)$ and the single-particle Green's function $G_\sigma(p)$, the
Bethe-Salpeter equation for the particle-particle channel is
\begin{equation}
\lambda_\alpha \phi_\alpha(p) =
-{T\over N} \sum_{p'} \Gamma_{\rm IS} (p|p') G_\uparrow(p')
G_\downarrow(-p') \phi_\alpha(p').
\label{eq:bs}
\end{equation}
Here, as before, the sum on $p'$ is over both ${\bf p}'$ and
$\omega_{n'} = (2\,n'+1)\pi T$.
Using the same Monte Carlo data for the four-point vertex, one can also
determine the irreducible particle-hole vertex 
$\overline{\Gamma}_{\rm I}(p|p')$ for a
center-of-mass momentum $\QQ=\pp$ and spin 1.
Then, using this as a kernel, the Bethe-Salpeter equation for the
$\QQ=\pp$ particle-hole channel is
\begin{equation}
\overline{\lambda}_\alpha \overline{\phi}_\alpha(p) =
-{T\over N} \sum_{p'} 
\overline{\Gamma}_{\rm I} (p|p') G_\uparrow(p'+Q)
G_\downarrow(p') 
\overline{\phi}_\alpha(p').
\label{eq:ph}
\end{equation}

\begin{figure}
\centerline{\epsfysize=6.8cm \epsffile[-30 184 544 598] {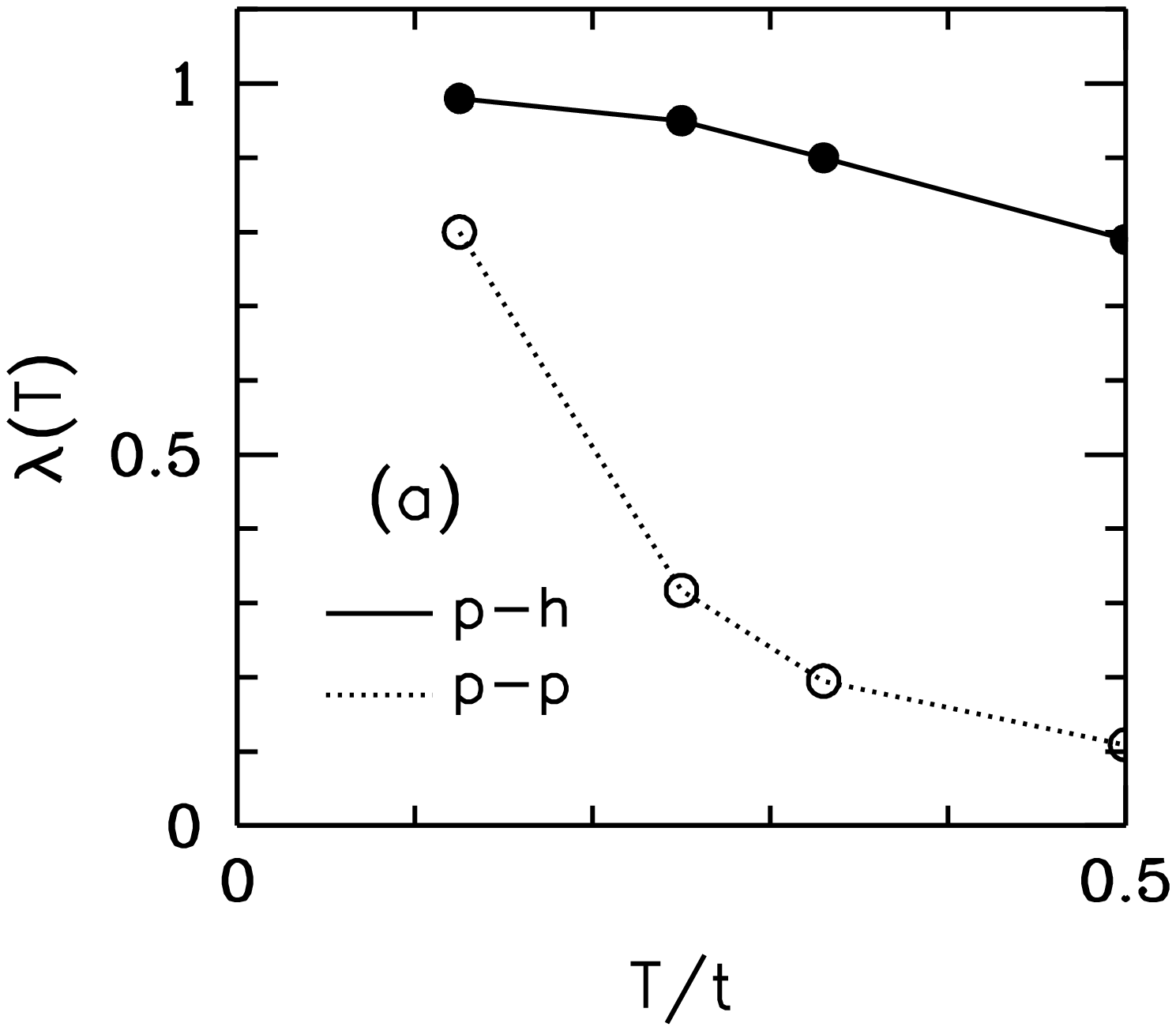}
\epsfysize=6.8cm \epsffile[98 184 672 598] {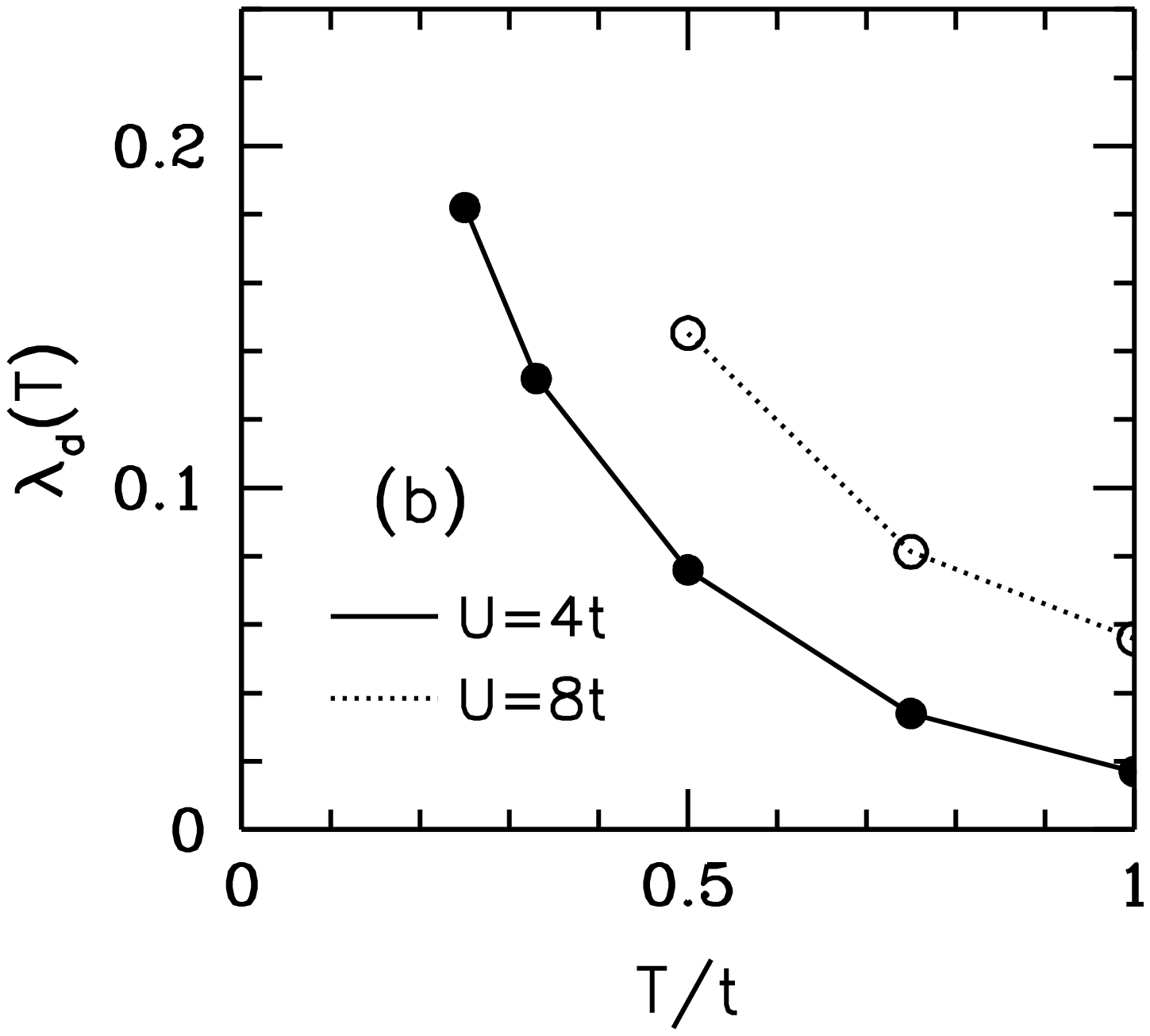}}
\vspace{0.3cm}
\caption{
(a) Leading $\QQ=\pp$, $S=1$ particle-hole (solid circle)
and the $\QQ=0$, singlet particle-particle (open circle) eigenvalues
versus temperature for $U/t=4$ and $\nang=1$.
(b) Singlet $d_{x^2-x^2}$ eigenvalue versus temperature for
$U/t=4$ (solid circle) and $U/t=8$ (open circle) at $\nang=0.875$.
\label{fig:lambda}}
\end{figure}


\begin{figure}
\centerline{\epsfysize=8cm \epsffile[18 184 592 598] {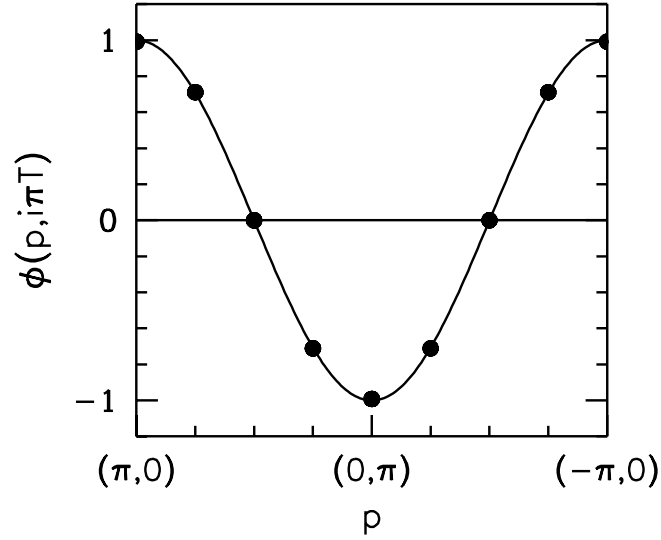}}
\vspace{0.3cm}
\caption{
Momentum dependence of the leading singlet even-frequency
particle-particle eigenfunction $\phi(\p,i\om_n)$ 
at $\om_n=\pi T$ for $U/t=4$, $\nang=0.875$ and $T=0.25t$.
Here, $\p$ moves on the fermi surface of the 
half--filled noninteracting system from $(\pi,0)$ to $(-\pi,0)$,
and $\phi(\p,i\pi T)$ has been normalized by its value at 
$\p=(\pi,0)$.
The solid line denotes
$(\cos{p_x}-\cos{p_y})/2$.  
Hence, near the Fermi surface $\phi(\p,i\pi T)$ is very close to the
$(\cos{p_x}-\cos{p_y})/2$ form.
\label{fig:phi}}
\end{figure}

\begin{figure} 
\centerline{\epsfysize=8cm \epsffile[18 184 592 598] {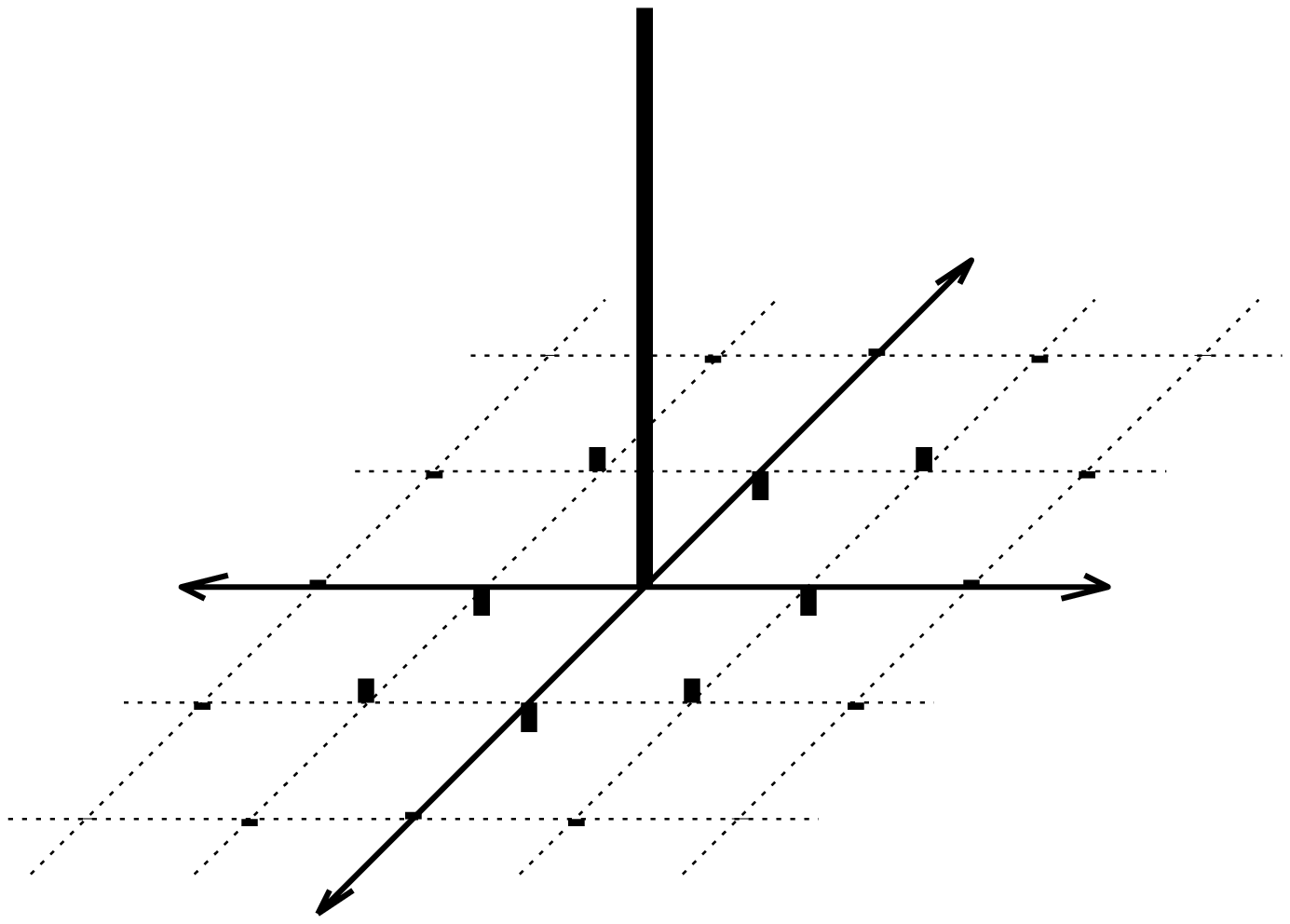}}
\vspace{-1.5cm}
\caption{
Real space structure of $\Gamma_{\rm IS}(\lb)$ for 
$U/t=4$, $\nang=0.875$ and $T=0.25t$.
\label{fig:vr}}
\end{figure}

Results showing the temperature dependence of the leading even frequency
eigenvalues for the particle-particle ($\QQ=0$) and particle-hole
($\QQ=\pp$) channels for a half-filled $8\times8$ lattice with $U/t=4$ are
plotted in Fig.~\ref{fig:lambda}(a).
At half-filling, the dominant eigenvalue occurs in the particle-hole
$\QQ=\pp$ spin $S=1$ channel, reflecting the development of long-range
antiferromagnetic correlations in the half-filled ground state.
The $\QQ=(\pi,\pi)$ magnetic susceptibility varies as 
\begin{equation}
\chi(T)\sim (1-\overline{\lambda}(T))^{-1}
\label{eq:chiT}
\end{equation}
and diverges as $T$ goes to zero.
The leading even frequency singlet particle--particle eigenvalue 
is associated with 
an eigenfunction $\phi(\p,i\om_n)$ which has $\dxy$ symmetry
as shown in Fig.~\ref{fig:phi}.
Its rapid increase at low temperature reflects the fact that the 
doped two--hole (or two--particle) state is bound~\cite{Dag,Par}.

For $\nang=0.875$, the Monte Carlo calculations are limited to 
higher temperatures because of the fermion determinantal 
sign problem.
However, as seen in Fig.~\ref{fig:lambda}(b) for $U/t=4$ and $8$,
the $\dxy$ eigenvalue increases as the temperature is lowered.
This is associated with the development of short range antiferromagnetic
correlations and the increase in $\Gamma_{\rm IS}$ at large 
momentum transfer. 
In this context it is interesting to plot the real space Fourier
transform
\begin{equation}
\Gamma_{\rm IS}(\lb) = 
{1\over N} \sum_{\q} {\rm e}^{i\q\cdot \lb}
\,\Gamma_{\rm IS}(\q,i\om_m=0)
\label{Gamell}
\end{equation}
As seen in Fig.~\ref{fig:vr}, $\Gamma_{\rm IS}(\lb)$
is repulsive on site but attractive on near neighbor sites.
It is this feature along with the large Fermi surface that leads to the 
$\dxy$ pairing fluctuations in the two--dimensional Hubbard model.

\section{Ladders}

New materials consisting of arrays of weakly coupled Cu--O ladders have
recently been synthesized~\cite{Tan}.
These systems provide further opportunities to examine the physical
properties of strongly correlated insulators and the metallic state
which is reached when holes are doped into such insulators.
Here we present a short review of the results which have been obtained
for the two-chain Hubbard ladder using a numerical density matrix
renormalization group (DMRG) method~\cite{Noack}.

\begin{figure} 
\begin{picture}(30,15)
\end{picture}
\includegraphics{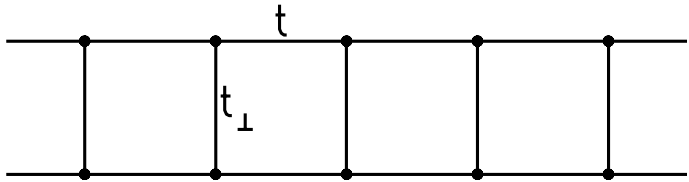}
\vspace{3.5cm}
\caption{
Two--chain Hubbard lattice with inter-chain and intra-chain hopping matrix
elements $t_{\perp}$ and $t$, respectively.
\label{fig:lattice}}
\end{figure}

\begin{figure}
\vspace{1cm}
\centerline{\epsfysize=7cm \epsffile[18 184 592 598] {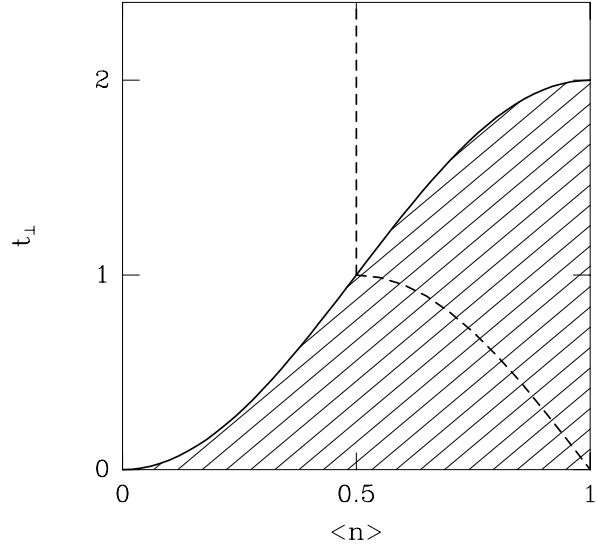}}
\vspace{0.3cm}
\caption{
Phase diagram of the $U=0$ system.
\label{fig:phase}}
\end{figure}

\begin{figure}
\centerline{\epsfysize=6cm \epsffile[-20 234 554 648] {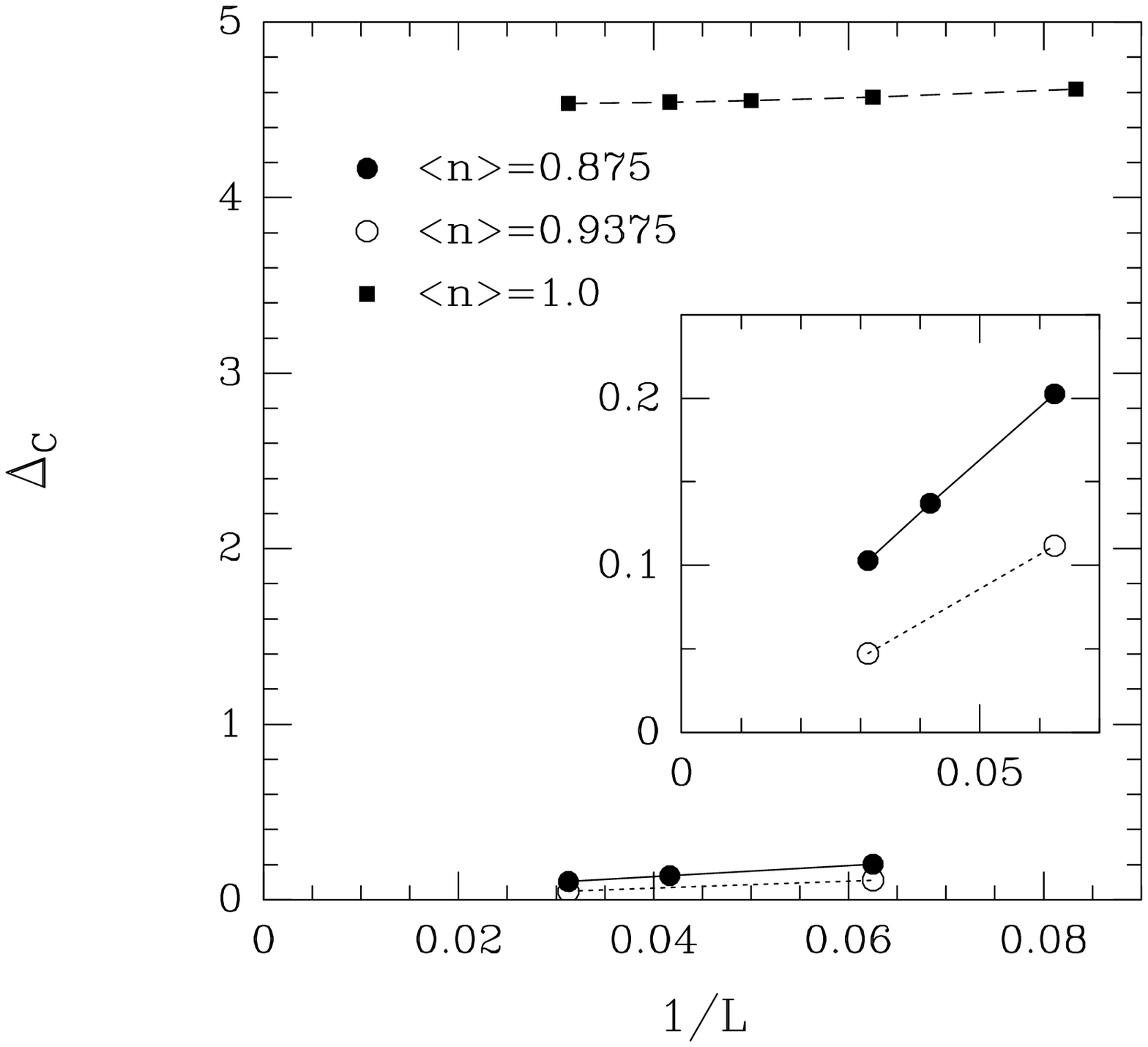}
\epsfysize=6cm \epsffile[28 234 602 648] {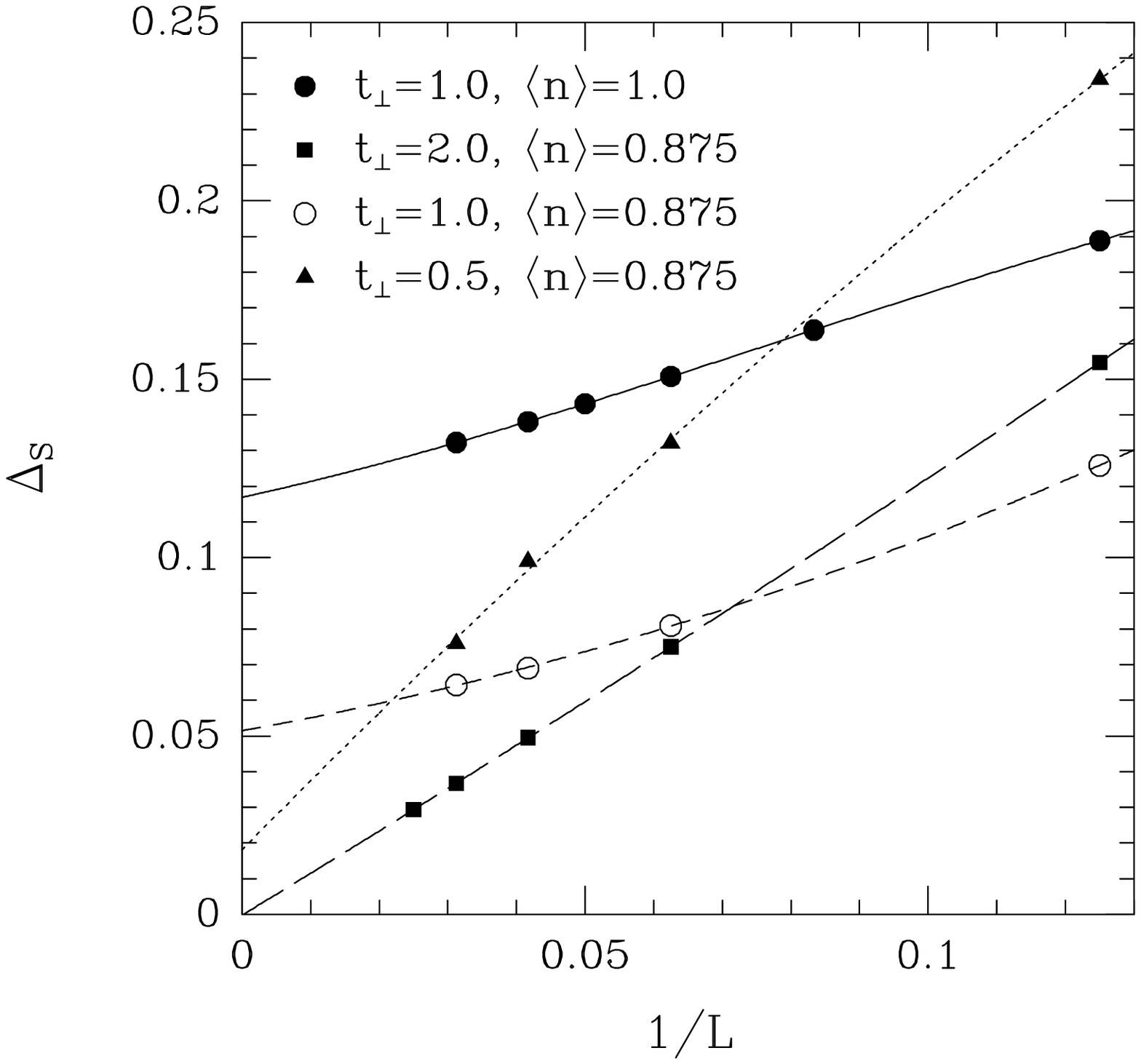}}
\vspace{1.1cm}
\caption{
(a) Charge and (b) spin gaps, $\Delta_c$ and $\Delta_s$,
versus the inverse lattice length $L^{-1}$
for $U/t=8$.
\label{fig:deltacs}}
\end{figure}

The two-chain Cu--O ladders in SrCuO$_3$ and La$_2$Cu$_2$O$_5$ consist
of Cu--O--Cu legs joined by placing an O atom between the Cu atoms of
the legs, forming Cu--O--Cu rungs.
In both cases, the ladders are also coupled to each other, but here we
consider the idealized case of an isolated ladder.
Furthermore, we approximate the ladder by the two-chain Hubbard lattice
shown in Fig.~\ref{fig:lattice}.
Here
\begin{equation}
H=-t \sum_{i,\lambda,\sigma} 
\left( c^\dagger_{i,\lambda,\sigma}c_{i+1,\lambda,\sigma} +
   c^\dagger_{i+1,\lambda,\sigma}c_{i,\lambda,\sigma} \right) -
t_\perp \sum_{i,\sigma}
  \left( c^\dagger_{i,1\sigma}c_{i,2\sigma} +
c^\dagger_{i,2\sigma}c_{i,1\sigma} \right)
 + U \sum_{i,\lambda} n_{i,\lambda\uparrow} 
n_{i,\lambda\downarrow}
\label{eq:hch}
\end{equation}
with $t$ the hopping integral parallel to the chains, $t_\perp$ the
hopping between the legs, and $U$ the onsite Coulomb interaction.
The operator $c^\dagger_{i,\lambda s}$ creates an electron of spin $s$
on site $(i,\lambda)$ with $\lambda = 1, 2$ denoting the leg and
$n_{i,\lambda s} = c^\dagger_{i,\lambda s} c_{i,\lambda s}$.
For $U=0$, $H$ can be diagonalized in terms of bonding ($-$) and
antibonding ($+$) bands with $\vep_k = 2\,t \cos k_x \pm t_\perp$.
A phase diagram for the two-chain ladder is sketched 
in Fig.~\ref{fig:phase} for $U=0$.
The region above the line $t_\perp/t = 1 - \cos(\pi\nang)$,
with $\nang = \langle n_{i\uparrow} + n_{i\downarrow} \rangle$ the site
filling, corresponds to having electrons only occupy the bonding band.
The shaded region has both the bonding and antibonding band occupied
with the bonding band half--occupied along the dashed curve.
Thus for the half-filled case, $\nang=1$, the bonding band is full when
$t_\perp/t$ exceeds 2 and the non-interacting system is a simple band
insulator with equal charge and spin gaps set by $2(t_\perp-2\,t)$.
In the non-interacting $U=0$ limit, for $t_\perp/t$ less than 2,
electrons occupy both bands and the system behaves as a simple metal
with vanishing charge and spin gaps.
Here we will see that the situation is quite different in the presence
of the Coulomb interaction $U$.
At half-filling with $U\neq0$, the system remains an insulator for
$t_\perp/t<2$, with a ratio of the spin gap to charge gap which
approaches zero smoothly as $t_\perp/t$ vanishes.
Furthermore, holes doped into this correlated insulator form $\dxy$-like
pairs which exhibit power-law correlations over much of the shaded
region, and while the charge-gap vanishes, a finite spin gap remains.

\begin{figure}
\centerline{\epsfysize=8cm \epsffile[18 184 592 598] {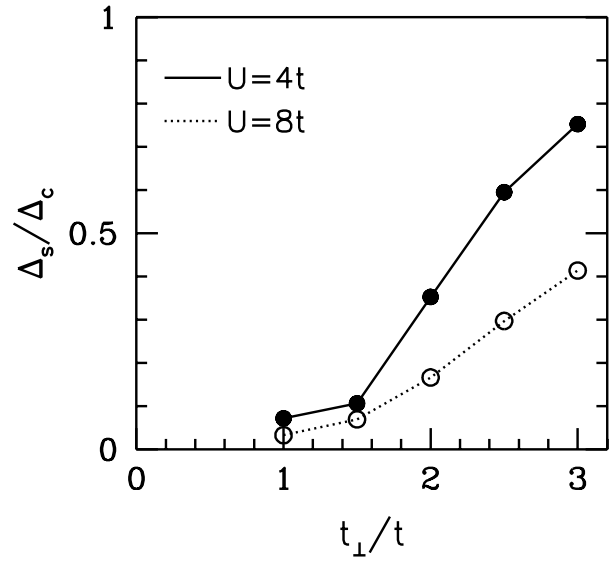}}
\vspace{0.6cm}
\caption{
Ratio of the spin and charge gaps, $\Delta_s/\Delta_c$,
versus $\tp/t$ for $U/t=4$ and 8
at $\nang=0.875$.
\label{fig:ratio}}
\end{figure}

The charge and spin gaps can be determined from the ground state energy 
$E_0(N_\uparrow,N_\downarrow)$ of a configuration with $N_\uparrow$ up-spin 
and $N_\downarrow$ down-spin electrons.
The charge gap $\Delta_c$ is
\begin{equation}
\Delta_c = {1\over 2} \left[ E_0(N_\uparrow + 1, N_\downarrow + 1) - 
       E_0(N_\uparrow,N_\downarrow) \right],
\label{eq:dc}
\end{equation}
and the spin gap
\begin{equation}
\Delta_s = E_0(N_\uparrow + 1, N_\downarrow - 1) -
E_0(N_\uparrow,N_\downarrow). 
\label{eq:ds}
\end{equation}
Here $N_\uparrow = N_\downarrow$ and $N_\uparrow + N_\downarrow$ is
fixed to give the desired filling $\nang = (N_\uparrow +
N_\downarrow)/L$
for a $2\times L$ site ladder.
Using the DMRG, calculations on ladders of varying lengths ($2\times L$)
have been carried out, giving the $L\to\infty$ extrapolations for
$\Delta_c$ and $\Delta_s$ shown in Fig.~\ref{fig:deltacs} 
for $U/t=8$ and various
fillings.
In this way, one finds that $\Delta_c$ and $\Delta_s$ are both finite
for $\nang=1$ but $\Delta_c$ vanishes for $\nang\neq1$.
A plot of $\Delta_s/\Delta_c$ versus $t_\perp/t$ for $\nang=1$ is shown
in Fig.~\ref{fig:ratio} for several values of $U/t$.
For the interacting system $\Delta_s/\Delta_c$ gradually approaches 1 as
$t_\perp/t$ increases beyond 2.
The difference ($1-\Delta_s/\Delta_c$) is one characteristic difference
between the strongly correlated insulating state and the band insulating
state.

\begin{figure}
\centerline{\epsfysize=8cm \epsffile[18 184 592 598] {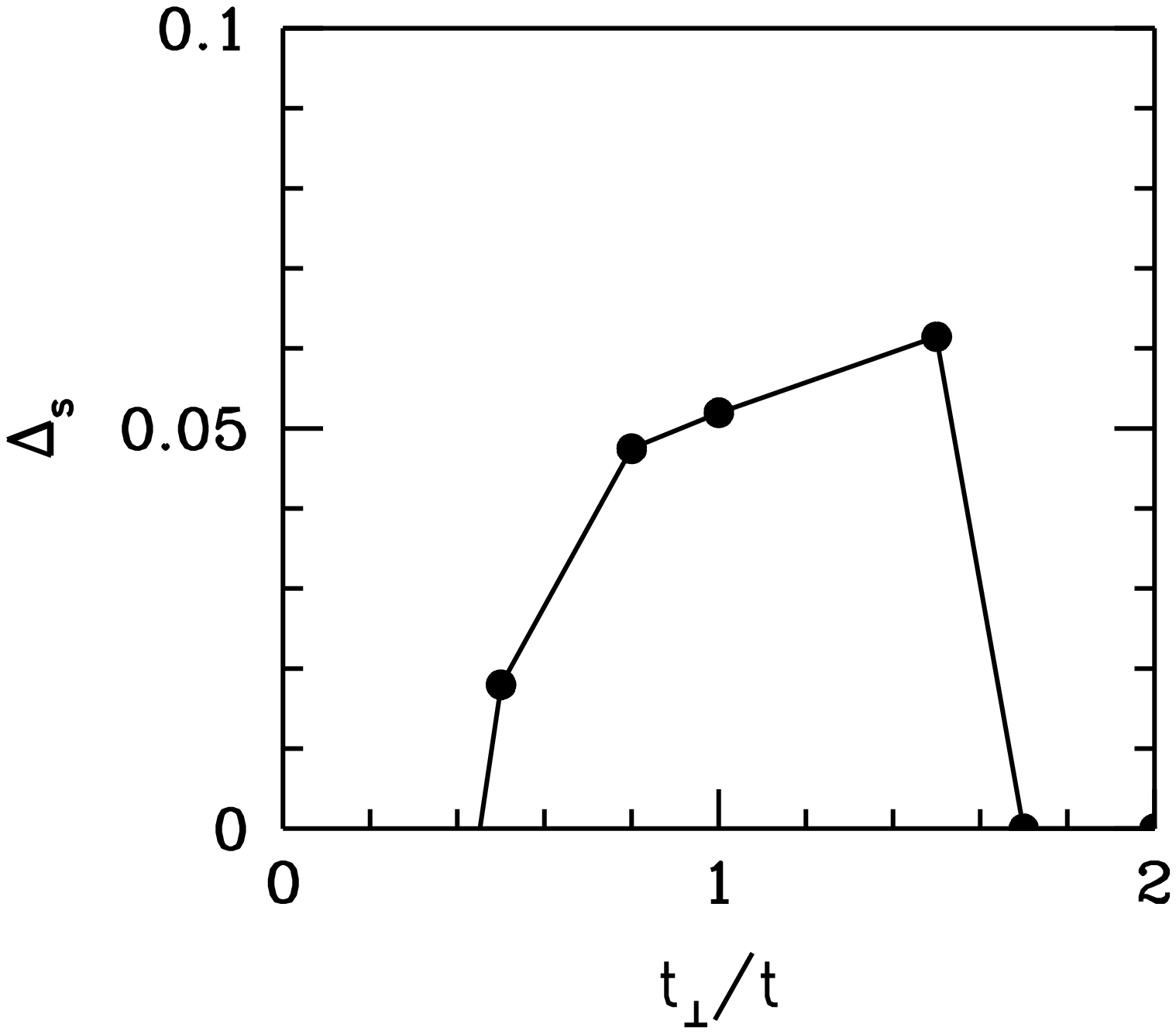}}
\vspace{0.6cm}
\caption{
Spin gap $\Delta_s$ versus $\tp/t$ for $U/t=8$ and $\nang=0.875$.
\label{fig:ds}}
\end{figure}

As noted, when holes are doped into the half-filled ladder, the charge
gap vanishes.
However, a spin gap remains which for a fixed value of $U/t$ depends
upon the ratio of $t_\perp/t$ and the filling.
In the following discussion, we will set $U/t=8$ and $\nang=0.875$,
and examine the properties of the system 
for various values of $t_\perp/t$. 
The dependence of the spin gap on $t_\perp/t$ is 
illustrated in Fig.~\ref{fig:ds}.
For $t_\perp/t \gtwid 1.7$, the spin gap vanishes, consistent with the
one-band Luttinger liquid picture suggested by Fig.~\ref{fig:phase}.
For $U=0$, the anti-bonding band becomes empty for $\nang=0.875$ when
$t_\perp/t$ exceeds 1.85.
As $t_\perp/t$ decreases, the spin gap rises and then appears to vanish
for $t_\perp/t \simeq 0.43$.
This is near the point at which the bonding band is half-filled, and as
discussed by Balents and Fisher~\cite{Bal}, 
a RNG calculation finds that the spin
gap vanishes.
At still lower values of $t_\perp/t$, the RNG calculation finds a small
spin gap, but we have been unable to resolve it if it is indeed
present.

\begin{figure}
\centerline{\epsfysize=8cm \epsffile[18 184 592 598] {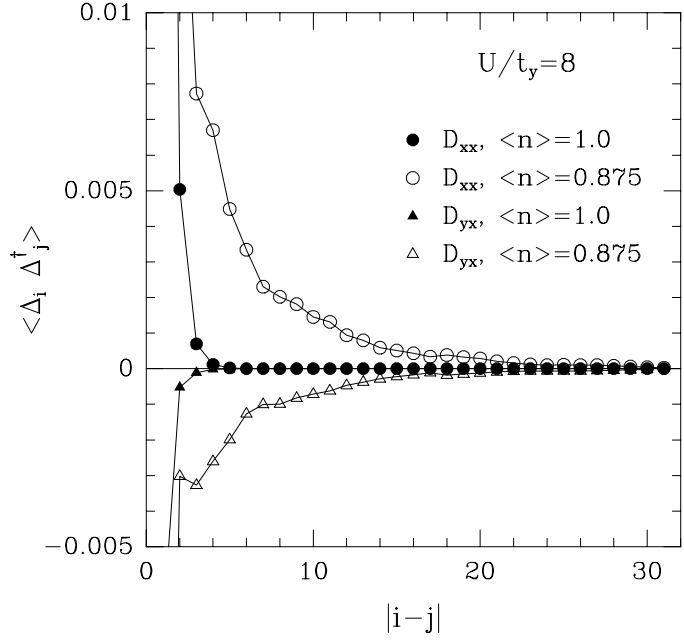}}
\caption{
Pair field correlation functions $D_{xx}(\ell)$ and $D_{yx}(\ell)$
versus the lattice spacing $\ell=|i-j|$.
\label{fig:dxx}}
\end{figure}

\begin{figure}
\centerline{\epsfysize=8cm \epsffile[18 184 592 598] {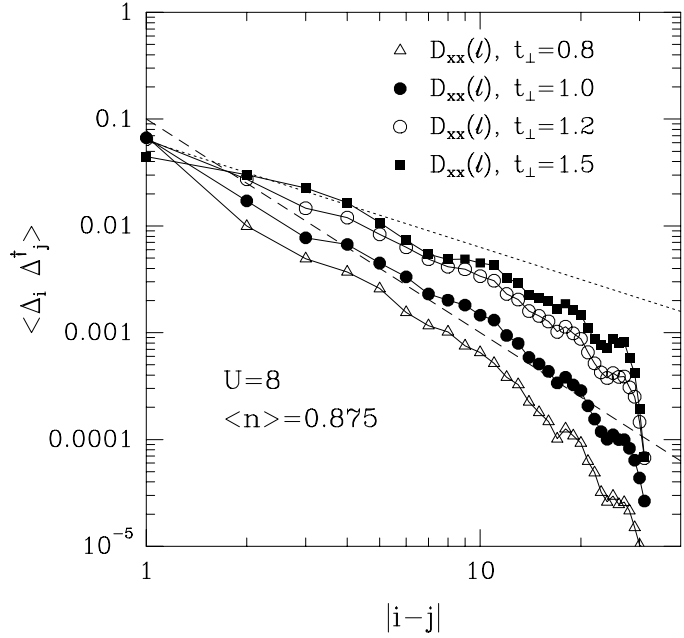}}
\caption{
Power law decay of $D_{xx}(\ell=|i-j|)$ for different values of
$\tp/t$ at $U/t=8$ and $\nang=0.875$.
The dotted and the dashed lines decay as 
$|i-j|^{-1}$ and $|i-j|^{-2}$, respectively.
\label{fig:power}}
\end{figure}

\begin{figure}
\centerline{\epsfysize=8cm \epsffile[18 184 592 598] {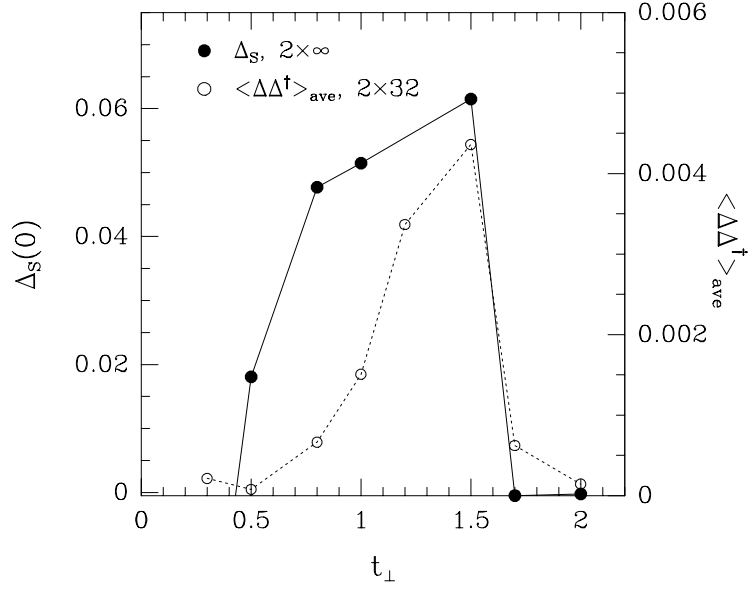}}
\caption{
Comparison of the strength of the pairing correlations (open circle)
with the
size of the spin gap $\Delta_s$ (solid circle)
for $U/t=8$ and $\nang=0.875$.
The average over $\ell$ of 
$\langle \Delta_i \Delta_{i+\ell} \rangle_{\rm ave}$
has been done for $\ell$ between 8 and 12 lattice spacings.
\label{fig:comp}}
\end{figure}

\begin{figure}
\centerline{\epsfysize=8cm \epsffile[18 184 592 598] {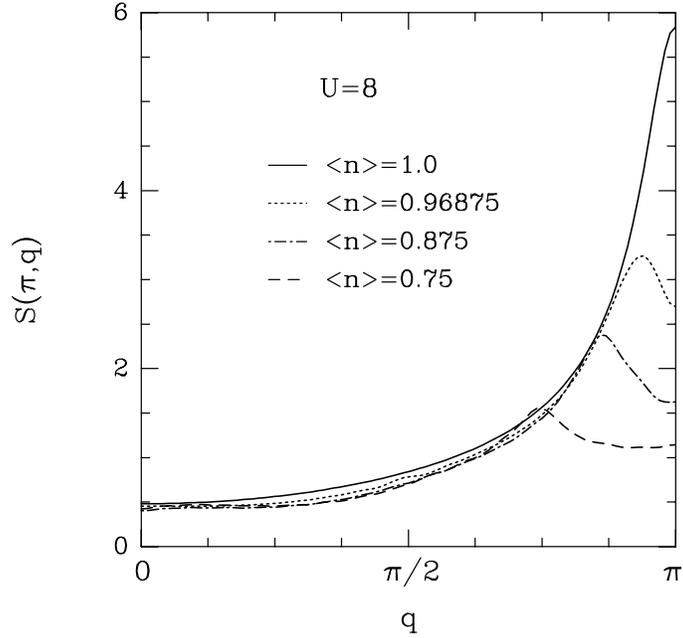}}
\caption{
Magnetic structure factor $S(\pi,q)$ versus $q$
along the ladder for $U/t=8$ and various fillings.
\label{fig:Sq}}
\end{figure}

In the spin gap regime of the doped system, the dominant two-particle
correlations appear in the pairing channel.
Figure \ref{fig:dxx} 
shows the rung-rung and rung-leg pair-field correlation
functions
\begin{equation}
D_{xx}(\ell) = 
\left\langle \Delta_x(i+\ell) \Delta^\dagger_x(i) \right\rangle
\label{eq:dxx}
\end{equation}
and
\begin{equation}
D_{yx}(\ell) = 
\left\langle \Delta_y (i+\ell) \Delta^\dagger_x(i)
\right\rangle.
\label{eq:dxy}
\end{equation}
Here
\begin{equation}
\Delta^{\dagger}_x(i) =  
\left( c^{\dagger}_{1i\uparrow}c^{\dagger}_{2i\downarrow} - 
c^{\dagger}_{1i\downarrow} c^{\dagger}_{2i\uparrow} \right)  
\label{eq:dx}
\end{equation}
creates a singlet pair on the $i$th rung, while
\begin{equation}
\Delta^{\dagger}_y(i) =  
\left( c^{\dagger}_{1i\uparrow}c^{\dagger}_{1i+1\downarrow} -
c^{\dagger}_{1i\downarrow}c^{\dagger}_{1i+1\downarrow} \right)  
\label{eq:dy}
\end{equation}
creates a singlet pair along a leg of the ladder at site $i$. 
The Cu--O hopping parameters $t$ and $t_\perp$ have the same sign, so
that the relative phases of the pair-field correlations shown in 
Fig.~\ref{fig:dxx}
are meaningful and imply that the bound pairs have $\dxy$-like symmetry,
as discussed by Rice {\it et al.}~\cite{Rice}
The power law decay of these pairing correlations is shown in 
Fig.~\ref{fig:power}.
It is also interesting to compare the strength of the pairing
correlations with the size of the spin gap.
This is done in Fig.~\ref{fig:comp}.

These results support a picture in which the doped state of the two-leg
Hubbard ladder, in the appropriate $t_\perp/t$ regime, is characterized
by a finite spin gap and $\dxy$-like power-law pairing correlations.
In Fig.~\ref{fig:Sq}, 
we show plots of the magnetic structure factor $S(\pi,q)$
with $q$ along the ladder for various fillings.
In the doped ladders, the peak in the structure factor occurs at an
incommensurate wavevector.
The exponential decay of the magnetization--magnetization correlation
functions in the spin--gapped regime lead to Lorentzian line shapes.
Thus even though the antiferromagnetic correlations are 
clearly short range, they lead to power law $d_{x^2-y^2}$ 
pairing correlations.

\section{Conclusions}

The development of both the low-energy quasiparticle dispersion and the
peak in the singlet particle-particle vertex at large momentum transfers
reflects the growth of the short-range antiferromagnetic correlations as
$T$ decreases below $J$.
Analysis of the Bethe-Salpeter equation shows that the leading
even-frequency singlet pairing occurs in the $\dxy$ channel.
Physically, for the large Fermi surface associated with the observed
quasiparticle dispersion, a particle-particle vertex which increases at
large momentum transfer favors $\dxy$ pairing. 
Note that the tendency for $\dxy$ pairing does not require a particularly 
sharp or narrow peak in
$\Gamma_{\rm IS}(p-p')$ for ${\bf p-p'} = \pp$, but rather simply
weight at large momentum transfers corresponding to strong short-range
antiferromagnetic correlations.
This is clearly seen in the results for the two-leg ladder
where the antiferromagnetic correlations exponentially decay.
Thus it is the strong short--range antiferromagnetic
correlations which lead to the formation of $d_{x^2-y^2}$
pairing correlations in the Hubbard model.

\section*{Acknowledgments}

The work on the 2D Hubbard model discussed here was carried out
with N. Bulut and S.R. White. 
The work on the ladder was carried out with R. Noack and S.R. White.
This work was supported in part by the National Science Foundation under
grant No.~DMR92--25027.
The numerical computations reported in this paper
were carried out at the San Diego Supercomputer Center.


\end{document}

%% file: defs.tex
\def\pmb#1{\setbox0=\hbox{#1}%
  \kern-.025em\copy0\kern-\wd0
  \kern.05em\copy0\kern-\wd0
  \kern-.025em\raise.0433em\box0 }

\def\nang{\langle n\rangle}

\def\vep{\varepsilon}
\def\dxy{d_{x^2-y^2}}

\def\lb{\pmb{$\ell$}}
\def\pp{(\pi,\pi)}

\def\txN{{\textstyle{1\over N}}}
\def\txN2{\textstyle{1\over N^2}}

\def\im{{\rm Im\,}}

\def\cuo2{CuO$_2$}

\def\2kf{2\,k_F}
\def\4kf{4\,k_F}

\def\tp{t_\perp}

\def\q{{\bf q}}
\def\p{{\bf p}}

\def\om{\omega}